\documentclass[a4paper,fleqn]{cas-sc}

\usepackage[authoryear,longnamesfirst]{natbib}
\usepackage[utf8]{inputenc}
\usepackage{graphicx}
\usepackage{xcolor}
\usepackage{soul}
\usepackage{booktabs}
\usepackage{algorithmic}
\usepackage[linesnumbered,ruled,vlined]{algorithm2e}
\usepackage{mathtools}
\usepackage{caption} 
\usepackage{courier}  
\usepackage{framed}
\usepackage{multirow}
\usepackage{enumitem}
\usepackage{tcolorbox} 
\usepackage{colortbl}
\usepackage{pgfplots}
\pgfplotsset{width=10cm,compat=1.9}
\usepackage{threeparttable}    
\usepackage{subcaption}
\usepackage[normalem]{ulem}
\definecolor{cGreen}{RGB}{0,150,0}
\definecolor{brown}{RGB}{139,64,0}
\newcommand{\gray}[1]{{\color{gray}~#1 }}

\usepackage{pgfkeys}
    \newenvironment{customlegend}[1][]{%
        \begingroup
        \csname pgfplots@init@cleared@structures\endcsname
        \pgfplotsset{#1}%
    }{%
        \csname pgfplots@createlegend\endcsname
        \endgroup
    }%
    \def\addlegendimage{\csname pgfplots@addlegendimage\endcsname}
\definecolor{myboxcolor2}{RGB}{191,128,191}
\newtcolorbox{mybox3}[1]{
  colback=myboxcolor2!5!white,
  colframe=myboxcolor2!75!black,
  fonttitle=\bfseries,
  title=#1,
  halign title=center, 
  halign=flush left,    
  left=1mm,
  right=1mm
}

\newcommand{\ournameAbbr}{{TF-DCon} }

\def\tsc#1{\csdef{#1}{\textsc{\lowercase{#1}}\xspace}}
\tsc{WGM}
\tsc{QE}
\tsc{EP}
\tsc{PMS}
\tsc{BEC}
\tsc{DE}

\begin{document}
\let\WriteBookmarks\relax
\def\floatpagepagefraction{1}
\def\textpagefraction{.001}

\shorttitle{Dataset Condensation for Content-based Recommendation}

\shortauthors{Jiahao Wu et~al.}

\title [mode = title]{Leveraging ChatGPT to Empower Training-free Dataset Condensation for Content-based Recommendation}                      



%
\author[1,2]{Jiahao Wu}


\fnmark[1]

\ead{notyour_mason@outlook.com}



\affiliation[1]{organization={Southern University of Science and
Technology},
    country={China}
    }
\affiliation[2]{organization={The Hong Kong Polytechnic University},
    country={Hong Kong}
    }
\affiliation[3]{organization={National University of Singapore},
    country={Singapore}}

\author[2]{Qijiong Liu}
\ead{liu@qijiong.work}
\fnmark[1]

\author[3]{Hengchang Hu}
\ead{holdenhhc@gmail.com}



\author[2]{Wenqi Fan}
\ead{wenqifan03@gmail.com}




\author[1]{Shengcai Liu}
\ead{liusc3@sustech.edu.cn}

\author[2]{Qing Li}
\ead{csqli@comp.polyu.edu.hk}

\author[2]{Xiao-Ming Wu}
\ead{xiao-ming.wu@polyu.edu.hk}

\author[1]{Ke Tang}
\ead{tangk3@sustech.edu.cn}

\fntext[fn1]{Equal contribution.}
\cortext[cor1]{Corresponding author}


\begin{abstract}
Modern techniques in Content-based Recommendation (CBR) leverage item content information to provide personalized services, effectively alleviating information overload. However, these methods suffer from resource-intensive training on large datasets. To address this issue, we explore the dataset condensation for textual CBR in this paper. Dataset condensation aims to synthesize a compact yet informative dataset that enables models to achieve performance comparable to those trained on full datasets.  
Directly applying existing approaches to CBR has limitations: (1) difficulties in synthesizing discrete texts and (2) the inability to retain preference information between users and items during condensation. To bridge this gap, we propose TF\_DCon, an efficient dataset condensation method for CBR. Leveraging ChatGPT, TF\_DCon enables condensation on discret texts and preserves preference via a clustering-based synthesis module to condense users. Unlike previous methods, TF\_DCon's forward pipeline eliminates the iterative updates on the condensed data, greatly enhancing the efficiency. Extensive experiments conducted on three real-world datasets demonstrate TF\_DCon's effectiveness. Notably, we are able to approximate up to $97\%$ of the original performance while reducing the dataset size by $95\%$ (i.e., on dataset MIND). 
\end{abstract}




\maketitle

\section{Introduction}
Content-based recommenders~\citep{mind2020acl,ijcai19naml,wu2021fastformer} have made strides in mitigating the information overload dilemma, delivering items with relevant content (i.e., news, articles, or movies) to users. Existing advanced content-based recommendation (CBR) models~\citep{dkn2018ww,nrms2019emnlp} are trained on 
{large-scale}
datasets encompassing millions of users and items. 
{Handling these large datasets significantly strains computational resources during model training. Furthermore, the recurrent retraining of recommendation models, especially for periodic updates in real-world applications, exponentially elevates costs to unsustainable levels.}


The intensifying exploration of \textit{dataset condensation}, also called \textit{dataset distillation}, shed light in addressing the aforementioned issues. The goal of dataset condensation is to synthesize a small yet informative dataset, trained on which the model can achieve comparable performance to that of a model trained on the original dataset~\citep{tongzhouw2018datasetdistillation,jinICLR2022graphCondesation,New2023GraphCondense,doscond-kdd2022,zhaoICLR2021DC,fair-text-dd-2022-emnlp}. The most prevalent paradigm~\citep{zhaoICLR2021DC,doscond-kdd2022,sucholutsky2021soft-text} is to formulate the condensation as bi-level optimization problem and iteratively update the condensed data by matching gradients of network parameters between synthetic and original data. Such methods have achieved success on condensing continuous data, such as images and node features.
However, condensing datasets for CBR is still unexplored.
\begin{figure}[]
\centering
{\includegraphics[width=0.95\linewidth]{{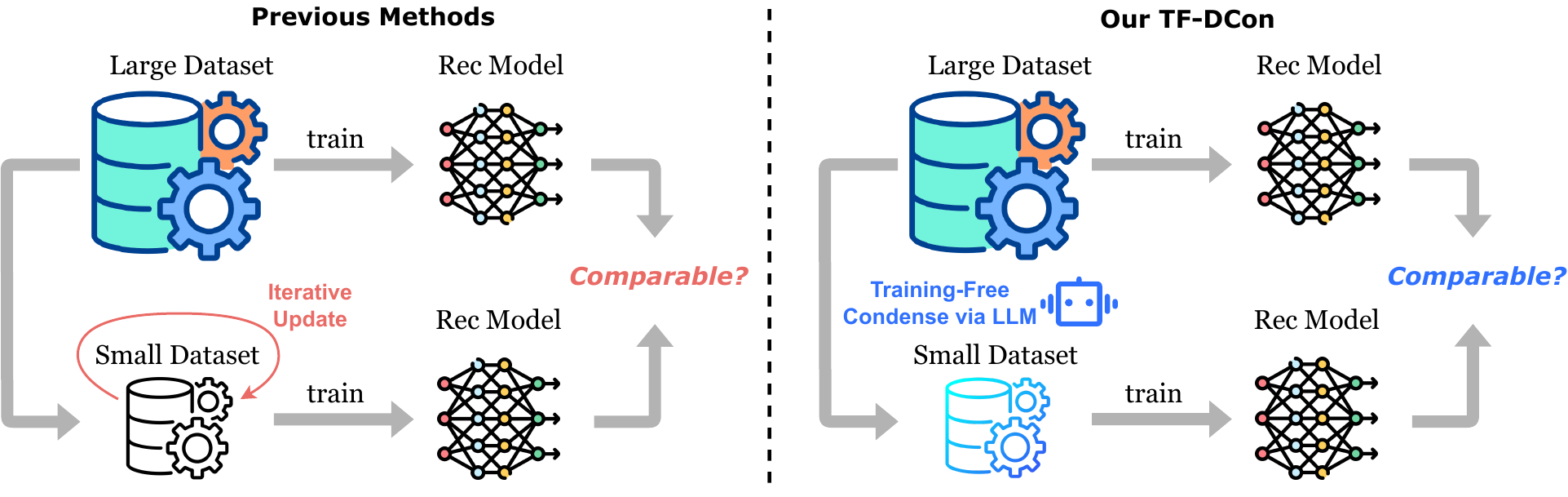}}}
\caption{Comparison on the pipelines of dataset synthesis between our \ournameAbbr  and previously proposed methods in other domains.}
\label{fig:intro-dd-compar}
\end{figure}


To bridge this gap, we investigate how to effectively condense the dataset for the textual content based recommendation. Existing condensation approaches devised in other domains~\citep{zhaoICLR2021DC,jinICLR2022graphCondesation,doscond-kdd2022,fair-text-dd-2022-emnlp} face two main challenges in the context of CBR: (1) \textit{How to generate discrete textual data?} Current condensation methods are designed for continuous data (i.e., images and text embeddings), following a formulation of nested bi-level optimization. Under this formulation, the data is synthesized via the outer gradient, which cannot be utilized to generate discrete textual data. Therefore, a solution for text synthesis is essential. (2) \textit{How to preserve the preference information between users and items?} In recommendation tasks, user and item data, along with their interactions, are critical for inferring user preferences. However, previously proposed methods mostly involve classification tasks, limiting their ability to handle interaction data and retain preference information.


Large Language Models (LLMs), such as ChatGPT developed by OpenAI, have shown a strong capacity to distill textual information~\citep{nips20LLM,arxiv2022LLM,jiatong2023empowering}, and its general expertise across various fields such as medicine~\citep{medicine1-2023GPT,medicine2-2023GPT}, recommendation~\citep{fan2023recommenderLLM,dai2023uncovering}, and law~\citep{jiaxi2023chatlaw,Jonathan2023chatgptLaw}.
With their versatility and extensive world knowledge, LLMs exhibit emergent abilities such as robust text comprehension and language generation~\citep{nips20LLM,arxiv2022LLM}. Leveraging these capabilities~\citep{nips20LLM,arxiv2022LLM}, we propose to leverage ChatGPT for dataset condensation, processing textual content to address one of the aforementioned challenges.

While LLMs are highly effective at processing text, they lack specific domain knowledge for recommendation scenarios and the ability to capture personalized user preferences. Naïvely applying LLMs for recommendation data condensation risks losing both this domain knowledge and essential preference information. 
To this end, we propose a ChatGPT-powered \textbf{T}raining-\textbf{F}ree \textbf{D}ataset \textbf{Con}densation method for content-based recommendation, abbreviated as \textbf{TF-DCon}. TF-DCon is devised in a two-level manner: \textit{content-level} and \textit{user-level}. 
At content-level, we curate a prompt-evolution module to optimize prompts, enabling ChatGPT to adapt to the specific recommendation domain and condense each item’s information into an informative title. At user-level, to capture the preference information of users, we propose a clustering-based synthesis module to simultaneously generate fake users and their corresponding historical interactions, based on user interests extracted by ChatGPT and user embeddings.
Consequently, our approach offers several advantages: (1) \ournameAbbr omits the conventional paradigm of formulating condensation as a bi-level optimization problem~\citep{zhaoICLR2021DC,tongzhouw2018datasetdistillation,jinICLR2022graphCondesation}. Instead, we formulate condensation as a forward process, eliminating iteratively training on datasets and greatly enhancing the condensation efficiency (see Figure~\ref{fig:intro-dd-compar}); (2) TF-DCon can generate text, which makes the condensed dataset more generalizable and flexible to train different architectures of models for CBR. (3) TF-DCon seamlessly embeds user preferences into the condensed data through the clustering-based synthesis module. Our contributions are summarized as follows:
\begin{itemize}[leftmargin=*]
    \item \textbf{Objective.} To the best of our knowledge, our study presents the first exploration of dataset condensation for textual CBR. We introduce a forward condensation paradigm distinct from traditional bi-level optimization and aim to effectively condense the information of items and user preference. 
    \item \textbf{Methodology.} We propose a ChatGPT-driven dataset condensation method featuring a prompt-evolution module for item content condensation and a clustering-based synthesis module to generate synthetic users and interactions, thereby preserving both item and preference information effectively.  
    \item \textbf{Experiment.} Extensive experimental results demonstrate the efficacy of the proposed method. In particular, we are able to approximate up to $97\%$ of the original performance while reducing the dataset size by $95\%$ (i.e., on the dataset MIND). The model training on the condensed datasets is significantly faster (i.e., $5\times$ speedup).
\end{itemize}
The rest of the paper is structured as follows. Section~\ref{sec:related-works} reviews the related works for content-based recommendation and dataset condensation. Section~\ref{sec:preliminaries} provides background for relevant preliminaries. Section~\ref{sec:methods} elaborates on the details of the proposed condensation method for CBR. Section~\ref{sec:experiments} presents the experimental results and analysis. Section~\ref{sec:limitations} discusses the limitations of our proposed method. The conclusion is obtained in section~\ref{sec:conclusion}.

\section{Related Work}
\label{sec:related-works}
\subsection{Content-Based Recommendation}

Content-based recommendation (CBR) involves a wide range of information in different formats: text~\citep{IPM2023YongqiLi,tkde2023wule-recSurvey}, image~\citep{IPM2024image1,IPM2024image2}, video~\citep{IPM2024video1,IPM2024video2}, social networks~\citep{IPM2024SocialSessionRec,IPM2024socialMultiModal}, etc. In this paper, we mainly focus on textual contents, i.e., news, books, and movie reviews.

In recent years, textual CBR has garnered substantial attention and has been the subject of comprehensive investigation within both industrial and academic domains. Models based on neural networks~\citep{dkn2018ww,nrms2019emnlp} have been proposed to learn the representations of users and items via CNN~\citep{CNN2012Nips} or attention mechanism~\citep{attention2017Nips}. Thereafter, the advance of pretrained language models like BERT~\citep{naacl-DevlinCLT19} exhibited promising performance on textual content recommendation~\citep{acl2022MINER,unbert2021ijcai}. As those methodologies depend on training with extensive-scale datasets, their exceptional performance comes at a high computational cost, which pose a challenge for sustainable model training and further scalable application. 
The recent emergence of investigations into dataset condensation and dataset distillation presents a promising avenue for tackling this concern~\citep{tongzhouw2018datasetdistillation,zhaoICLR2021DC}. The inherent concept underlying dataset condensation and dataset distillation revolves around the synthesis of a compact dataset for model training, with the objective of enabling models trained on this condensed dataset to achieve performance levels comparable to those trained on the complete dataset. Nevertheless, no extant research has delved into the realm of condensation specifically within the context of CBR. To bridge this critical gap, we introduce an innovative and efficacious framework tailored to the task of dataset condensation for CBR.
\subsection{Dataset Condensation}
It is notorious that training neural networks on large datasets is prohibitively expensive. To mitigate this issue, plenty of works on dataset distillation/dataset condensation have been proposed to synthesize small datasets to replace the large ones, such as GCond~\citep{jinICLR2022graphCondesation}, DC~\citep{zhaoICLR2021DC} and SFGC~\citep{New2023GraphCondense}. Those methods always formulate the condensation process as a bi-level optimization problem. The condensation methods have been explored on images~\citep{zhaoICLR2021DC}, graph~\citep{jinICLR2022graphCondesation} and text embeddings~\citep{fair-text-dd-2022-emnlp}. Various methods have been proposed to improve the quality of condensation and optimization efficiency. For instance, DC~\citep{zhaoICLR2021DC} proposed gradient matching, achieving a comparable performance under a high compression ratio, and DosCond~\citep{doscond-kdd2022} proposed one-step update strategy, which significantly improves the optimization efficiency. Condensation on textual data is conducted based on the text embeddings or soft labels~\citep{aclShort2023textdd-label-bert-finetune,lyq2021sigir-text-dd,multilingual-DD2023text,sucholutsky2021soft-text,fair-text-dd-2022-emnlp,Recsys2023KnowledgeFree}. Further exploration extends condensation methods to image-text retrival~\citep{wu2023multimodalDD}. Recently, a work~\citep{nips22infinite} aims at condensing collaborative filtering dataset but it is only suitable for NTK-based (Neural Tangent Kernel) architecture and fails to process textual data. Besides, CGM~\citep{ctrDC2023RecSys} follows the nested bi-level formulation to condense categorical data for CTR predictions and  DConRec~\citep{jiahao2023DConRec} designs an optimization-based method for ID-based recommendation, both of which cannot generate texts as well.

Despite the extensive investigation into dataset condensation, little attention has been paid to the domain of CBR. To address this disparity, we design a condensation method for the CBR dataset. Unlike earlier methodologies, our dataset condensation approach is devoid of training requirements, leading to a substantial reduction in condensation expenses. Furthermore, our condensation procedure centers on textual content, thereby diverging from those text condensation methods that hinge on text embeddings.
\section{Preliminaries}
\label{sec:preliminaries}

\begin{figure*}[]
\centering
{\includegraphics[width=0.995\linewidth]{{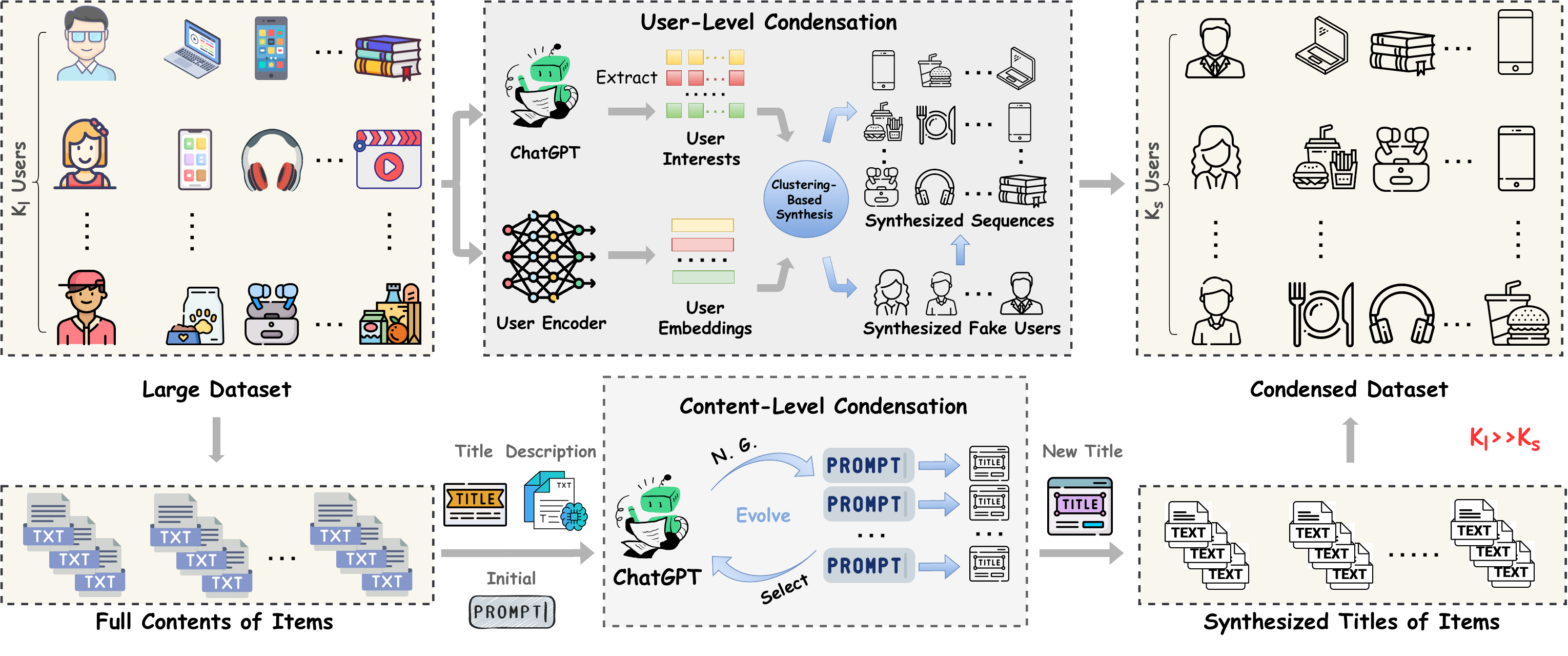}}}
\caption{The Proposed Method in a Nutshell. The outline of this method is presented in Algorithm~\ref{alg:tf-dcon} and the details of the prompt's evolution are outlined in Algorithm~\ref{alg:EvoPro}. ``N. G.'' denotes ``Next Generated Prompt Candidates''. Given a prompt, the content-level condensation and user interest extraction are instanced in Figure~\ref{fig:instances}.}
\label{fig:framework-dd}
\end{figure*}

\subsection{Content-based Recommendation}
Before elaborating on our proposed method, let us introduce some basic knowledge and a formal definition of content-based recommendation. CBR aims to infer the preference of users based on historical interactions and then recommend the contents with potential interest. In the context of CBR, the dataset $\mathcal{D}$ mainly consists of item set $\mathcal{N}$, user set $\mathcal{U}$, and click history set $\mathcal{H}$. Each item $n\in\mathcal{N}$ contains various contents, such as title, abstract, and category. Each user $u\in \mathcal{U}$ is associated with a clicking history of items $h^{(u)}\subset\mathcal{N}$. We denote the set of history $h^{(u)}$ by $\mathcal{H}$. Let $\mathcal{C}$ denote the set of clicks, where each click $c\in \mathcal{C}$ is defined as a tuple $(u,n)$ indicating that user $u$ has clicked on item $n$. The task of CBR is to infer the preference of users on given candidate items based on their historical interactions.

\subsection{Dataset Condensation}
 {The aim of dataset condensation} is to synthesize a small yet informative dataset $\mathcal{S}$,
trained on which the recommendation model can achieve comparable performance to the model trained on full dataset $\mathcal{D}$. Here, the synthesized dataset $\mathcal{S}$ also consists of its corresponding item set $\mathcal{N^S}$, user set $\mathcal{U^S}$ and click history set $\mathcal{H^S}$. Formally, the objective of the condensation could be formulated as follows:
\begin{equation}
    \label{eq:rec-data-condense-preliminary}
    \min_{\mathcal{S}} \mathcal{L}(f_{\boldsymbol{\theta}^\mathcal{S}}(\mathcal{D})), \text { s.t. } \boldsymbol{\theta}^\mathcal{S}=\underset{\boldsymbol{\theta}}{\arg\min\;}\mathcal{L}(f_{\boldsymbol{\theta}}(\mathcal{S})),
\end{equation}
where $\boldsymbol{\theta}$ is the parameter of recommendation model $f$ and $\mathcal{L}$ is the loss function. As we can see in Equation~\ref{eq:rec-data-condense-preliminary}, the condensation is a bi-level problem, where the outer optimization is to synthesize the condensed dataset and the inner optimization is to train the model on dataset $\mathcal{S}$.

\begin{algorithm}[]\small
    \caption{{TF-DCon}}\label{alg:tf-dcon}
    \begin{algorithmic}
        \STATE {\bfseries {Require:}} {dataset $\mathcal{D}=(\mathcal{H},\mathcal{N})$ where $\mathcal{H}$ is the set of users' historical interactions and $\mathcal{N}$ is the set of items, number of synthesized users in the condensed dataset $K$.}
    \end{algorithmic}
    \begin{algorithmic}[1]
        \STATE {Utilize ChatGPT to extract user interests: $\mathcal{I}\leftarrow ChatGPT(\mathcal{H},\mathcal{N})$.}
        \STATE {Encode user interests and user histories:$H\leftarrow UserEncoder(\mathcal{D})$, $H'\leftarrow TextEncoder(\mathcal{I})$.}
        \STATE {Clustering to generate $K$ synthetic users and their interacted item lists: $\mathcal{H}^{con}\leftarrow Clustering(K,H,H')$. //\gray{Section~\ref{subsec:user-level-condense}.}} 
        \STATE {Utilize ChatGPT to condense the content of items: $\mathcal{N}^{con}\leftarrow ChatGPT(\mathcal{N})$. //\gray{Section~\ref{subsec:news-level-condense}.}}
    \end{algorithmic}
    \begin{algorithmic}
        \STATE {\bfseries {Return}} {Condensed dataset $\mathcal{D}^{con}=(\mathcal{H}^{cond},\mathcal{N}^{cond})$.}
    \end{algorithmic}
\end{algorithm}
\section{{Method}}
\label{sec:methods}


In this section, we detail the proposed method (TF-DCon) and the overview is shown in Figure~\ref{fig:framework-dd}. Content-level condensation reduces the textual data load for each item (Section~\ref{subsec:news-level-condense}), while user-level condensation synthesizes fake users and interactions, capturing primary interests and merging users with similar preferences (Section~\ref{subsec:user-level-condense}). Section~\ref{subsec:discussion} compares the computational complexity of different condensation methods.

\subsection{{Content-Level Condensation}}
\label{subsec:news-level-condense}
\subsubsection{Content Condensation}
In the content-level condensation, we aim to condense all the information of each item into a succinct yet informative title. Recent studies~\citep{dai2023auggpt,ubani2023zeroshotdataaug} reveal the exceptional power of large language models in processing textual contents. Therefore, we propose to utilize ChatGPT to condense the contents. Specifically, we design the guiding prompt in the following format:
\begin{equation*}
    \begin{tabular}{l}
         {\textbf{Hints on the format of input:} }\\
     \textit{[title]\{title\}, [abs]\{abs\},...}\\

    {\textbf{Instructions on the format of output:}}\\
    
        \textit{[new\_title]\{new\_title\}}
    \end{tabular}\\
\end{equation*}

\noindent After facilitating ChatGPT's comprehension on the input, the information of items will be fed into ChatGPT to generate the condensed title. Formally, the process can be described as follows:
\begin{equation}
    n_s = ChatGPT(n),
\end{equation}
where $n\in \mathcal{N}$ is the original contents of item, consisting of $[title]$, $[abstract]$ and $[category]$, and $n_s\in \mathcal{N}^\mathcal{S}$ is the condensed title of the item, which only contain the $[title]$. An instance of the content-level condensation is depicted in Figure~\ref{fig:instances}.
\subsubsection{Prompt Evolution}
To efficiently guide ChatGPT to adapt to the recommendation scenario and perform thorough content condensation for each item, we propose a curated prompt evolution method, outlined in Algorithm~\ref{alg:EvoPro}. {Given an initial prompt, EvoPro can guide the prompts to evolve for pre-defined times. We name the initial prompt as \textit{0-th generation prompt}, abbreviated as \textit{$gen_0$-prompt}. During $i$-th iteration of evolving, EvoPro will first instruct ChatGPT to generate $N$ next-generation prompts $\{prompt_n\}$ based on \textit{$gen_{i-1}$-prompt}. Then, these prompts will be utilized to instruct ChatGPT to condense item contents for TF-DCon. The similarity scores between the embeddings of condensed contents and the original contents will be assigned to each prompt in  $\{prompt_n\}$, the calculation of which is outlined in Algorithm~\ref{alg:cal-score}. The prompt with the highest score will be selected as \textit{$gen_{i}$-prompt} for $i$-th generation. When the evolution finished, the prompt with the hightest score in the latest generation will be selected as the prompt for content condensation.

The rationale behind the score for prompt selection is that the condensed title sharing the largest similarity with the original contents preserves the most information. 
\begin{algorithm}[ht]\small
    \caption{{EvoPro}}\label{alg:EvoPro}
    \begin{algorithmic}
        \STATE {\bfseries {Require:}} {initial prompt \textit{$gen_0$-prompt}, item contents $\mathcal{C}=\{content_i\}$, evolving times $E$, number of prompts each generation $N$.}
    \end{algorithmic}
    \begin{algorithmic}[1]
        \STATE {Initial prompts set $\mathcal{P}\leftarrow$\{\} and counter $e\leftarrow0$ }
        \WHILE{{$e<E$}}
        \STATE {$e\leftarrow e+1$.}
        \STATE {Use ChatGPT to generate \textit{$gen_{e-1}$-prompt}'s $N$ next generation prompts $\{prompt_n\}$. }
        \STATE {Initialize the score set $\mathcal{S}\leftarrow\{\}$.}
        \FORALL{{$prompt$ in $\{prompt_n\}$}}
            \STATE {Calculate score for $prompt$: $s\leftarrow CalScore(prompt,\mathcal{C})$.}//\gray{Algorithm~\ref{alg:cal-score}.}
            \STATE {$\mathcal{S}\leftarrow\mathcal{S}\cup \{s\}$}
        \ENDFOR
        \STATE {Select the prompt with the highest score in $\mathcal{S}$ and set it as \textit{$gen_e$-prompt}.}
        \STATE {$\mathcal{P}\leftarrow\mathcal{P}\cup \{gen_e\text{-}prompt\}$.}
        \ENDWHILE
    \end{algorithmic}
    \begin{algorithmic}
       \STATE {\bfseries {Return}} {evolved prompts $\mathcal{P}$.}
    \end{algorithmic}
\end{algorithm}
\begin{algorithm}[htb]\small
    \caption{{CalScore}}\label{alg:cal-score}
    \begin{algorithmic}
        \STATE {\bfseries {Require:}} {$prompt$, item contents $\mathcal{C}$.}
    \end{algorithmic}
    \begin{algorithmic}[1]
        \STATE {$s\leftarrow 0$. }
        \FORALL{{$c_i$ in $\mathcal{C}$}}
            \STATE {Condense item content: $con_i\leftarrow LLM(prompt,c_i)$.}
            \STATE {$h_i\leftarrow TextEncoder(c_i),h'_i\leftarrow TextEncoder(con_i)$.}
            \STATE {Calculate the similarity score: $s\leftarrow s+Sim(h_i,h'_i)$.}
        \ENDFOR
    \end{algorithmic}
    \begin{algorithmic}
        \STATE {\bfseries {Return}} {$s$}
    \end{algorithmic}
\end{algorithm}

\subsection{User-Level Condensation}
\label{subsec:user-level-condense}
In user-level condensation, we aim to condense the number of users, reducing the total size of the interactions. First, we introduce the interest extraction for each user and user encoder. Then, we propose the {clustering-based synthesis} module to synthesize fake users and their corresponding historical sequences. In terms of user synthesis, we apply the clustering method to the user embeddings and each cluster corresponds to a synthetic user. To synthesize interactions for each user, we merge the interactions of users, whose embedding distance and interest distance to their cluster centroids rank top-$m$ while in descending order. 
\subsubsection{Interests Extraction and User Encoder} 
\textbf{{Interests Extraction:}} Each user $u$ is associated with a click history $h^{(u)}$, based on which we will utilize ChatGPT to extract the set of interests $\mathcal{I}^{(u)}$  for this user. Specifically, we design the guiding prompt in the following format:
\begin{equation*}
        \begin{tabular}{l}
         {\textbf{Hints on the format of input:} }\\
     \textit{(1)\{title\},  (2)\{title\},  (3)\{title\},  ...}\\

    {\textbf{Instructions on the format of output:}}\\
    
        \textit{[interests] -interest1,  -interest2,  ...}
    \end{tabular}
\end{equation*}
Then, the click history $h^{(u)}$ will be fed into ChatGPT to generate the interests of user $u$, which could be formulated as follows:
\begin{equation}
    \mathcal{I}^{(u)} = ChatGPT(h^{(u)}),
\end{equation}
where $\mathcal{I}^{(u)}$ is the set of interests for user $u$.

\textbf{{User Encoder:}} For each user $u$, given the associated click history $h^{(u)}$, we first we obtain the user's embedding as follows:
\begin{equation}
    \label{eq:user-embedding}
    z_u = f_{\theta}(h^{(u)}, \mathcal{N}),
\end{equation}
where $\mathcal{N}$ is the set of items and $f_{\theta}$ is the well-trained user encoder~\citep{nrms2019emnlp,ijcai19naml,wu2021fastformer}.
\subsubsection{Clustering-Based Synthesis} 
\textbf{{User Synthesis:}} Given the users' embeddings, we apply $K$-means algorithm over all of them to obtain their cluster centroids $\{c_i\}_{i=1}^{K_s}$, where $K_s$ is the number of synthesized users in the condensed dataset. 

\textbf{{Historical Sequence Synthesis:}} Given that each cluster corresponds to a fake user $u_s$, we need to synthesize the corresponding historical interactions. We devise a scoring mechanism based on the distance between user interests and user embeddings to select the historical interactions for fake users.

First, we calculate the distance between the embedding of each real user $u$ and its corresponding prototype $c_i$ as follows:
\begin{equation}
    \label{eq:dis-user-embedd}
    d^{emb}_u = Dis(z_u,c_i),
\end{equation}
where $Dis(\cdot,\cdot)$ is the distance function.

Given a set of interests $\mathcal{I}^{(u)}$ for each user $u$, we encode those interests by pretrained language model (PLM) as follows:
\begin{equation}
    \label{eq:plm-interest}
    e_u = f_{pool}\left(f_{PLM}(\mathcal{I}^{(u)})\right),
\end{equation}
where $f_{pool}$ is the pooling function and $f_{PLM}(\cdot)$ is the pretrained language model. Based on the clusters of user embeddings, we calculate the corresponding cluster centroids of user interests $\{c'_i\}_{i=1}^{K_s}$ by averaging the user interests in each cluster. Given the interest embeddings and their centroids, we can calculate the distance between them as follows:
\begin{equation}
    \label{eq:dis-user-interest}
    d^{int}_u = Dis(e_u,c'_i).
\end{equation}

\noindent Combining Equation~\ref{eq:dis-user-embedd} and Equation~\ref{eq:dis-user-interest}, we have:
\begin{equation}
    \label{eq:dis-user}
    d_u = d^{emb}_u + \alpha\cdot d^{int}_u,
\end{equation}
where $\alpha$ is a hyperparameter and $d_u$ is defined as the \textit{selection score}. Within each cluster, we employ an ascending ordering of users based on their selection scores. We subsequently curate the historical interactions for a synthetic user by merging the interaction histories of the top-$m$ users, which could be formalized as follows:
\begin{equation}
    \label{eq:history}
    h_s^{(u_s)} = \cup_{i=1}^{m} h^{(u_i)},
\end{equation}
where $d_{u_i}$ ranks top-$m$ within its corresponding cluster and $h^{(u)}\in \mathcal{H}$ is the historical interactions of $u$ in the original dataset.
\begin{figure*}[]
\centering
\includegraphics[width=1.0\columnwidth]{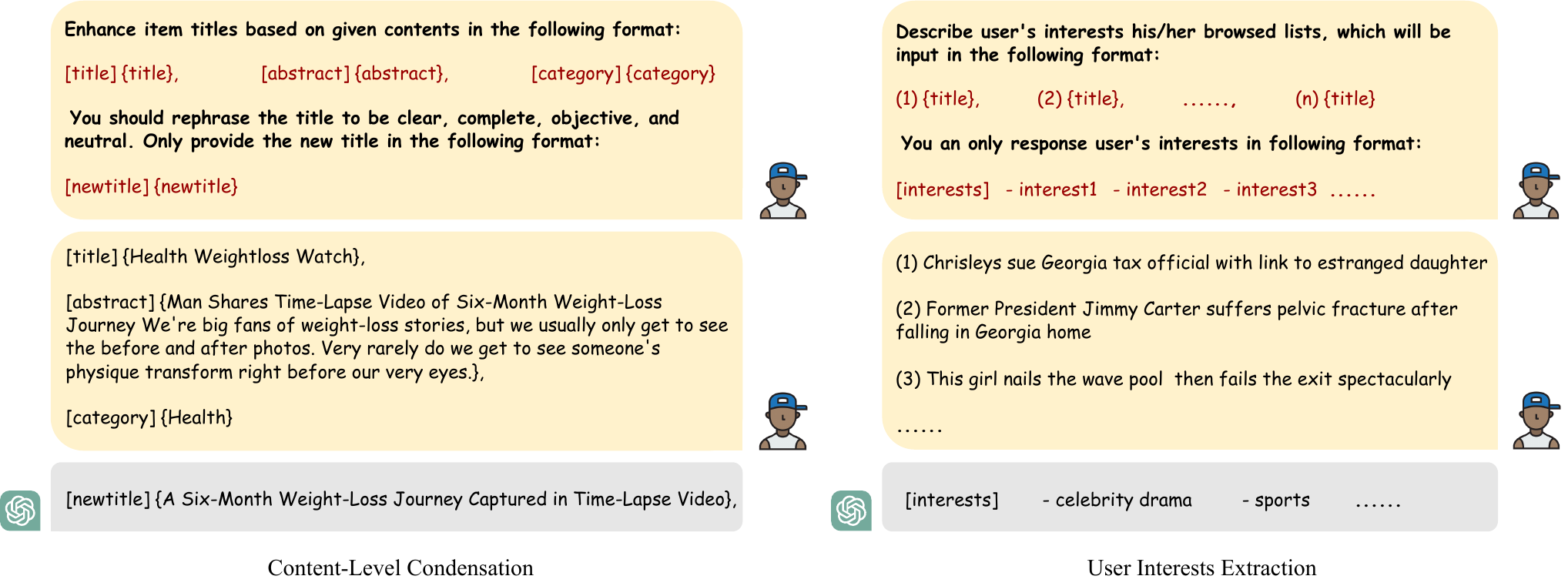}
\caption{An instance of content-level condensation and user interest extraction.}
\label{fig:instances}
\end{figure*}

\subsection{Computational Complexity Comparison.}

\begin{table}[]
\caption{Comparison of Computational Complexity.}
\begin{tabular}{l|l|lll}
\toprule
Category                        & Methods                                                & Inner    & Outer    & Total                                \\ \midrule
Kernel-based                    & Distill-CF~\citep{nips22infinite} & 1        & $\tau_o$ & $\tau_o$                             \\ \midrule
\multirow{2}{*}{Matching-Based} & DC~\citep{zhaoICLR2021DC}         & $\tau_i$ & $\tau_o$ & $\tau_i\cdot\tau_o$ \\ \cmidrule{2-5} 
                                & Doscond~\citep{doscond-kdd2022}   & 1        & $\tau_o$ & $\tau_o$                             \\ \midrule
Training-Free                   & TF-DCon (Ours)                                         & 1        & 1        & 1                                    \\ 
\bottomrule
\end{tabular}
\label{tab:compare-complexity}
\end{table}

\label{subsec:discussion}
{In this section, we discuss about the time complexity of TF-DCon and previous dataset condensation methods. The complexity in Table~\ref{tab:compare-complexity} is calculated by counting the times of generation/training based on the whole dataset.} Previous condensation methods are categorized into two types~\citep{pmlr2023ScaleUpDD}: matching-based and kernel-based. The first type generates synthetic datasets via gradient matchings between the
model trained on small-condensed dataset and those trained on the large-real dataset~\citep{zhaoICLR2021DC,doscond-kdd2022}, which typically involve a complex bi-level optimization. In the inner optimization, the models are trained while the datasets are optimized via outer gradients. In contrast, the kernel-based methods reduce the problem into a single-level optimization, which is always free from model training during condensation. If we denote the number of outer-iteration as $\tau_o$ and the number of inner-iteration as $\tau_i$, we can respectively obtain the computational complexity for each and we compare our proposed method with several representative condensation approaches. As shown in Table~\ref{tab:compare-complexity}, we can find that our proposed method is the most efficient since it follows a forward paradigm, thus free from iterative optimization on the synthesized dataset. 


\section{{Experiments}}
\begin{table}[]
\caption{Dataset statistics. ``avg. tok.'' denotes the average number of tokens in the content of each item and ``avg. his.'' denotes the average length of user histories.}
\centering
\begin{tabular}{l|l|l|l}
\toprule
{Datasets} & {MIND} & {GoodReads} & {MovieLens} \\ \midrule
\#{Items} & 65,238 & 16,833  & 1,682     \\ 
{avg. tok.}
& 45.42 & 35.37 & 34.80   \\ 
\midrule
\#{Users} & 94,057 & 23,089  & 943       \\
{avg. his.} & 14.98 & 7.81 & 4.94     \\ \midrule
\#{pos} & 347,727 & 273,888 & 48,884     \\
\#{neg} & 8,236,715 & 485,233 & 17,431   \\
{Density} & 0.0057\% & 0.0705\% & 3.0820\% \\
\bottomrule
\end{tabular}
\label{tab:dataset-statistics}
\end{table}

\label{sec:experiments}
In this section, we conduct experiments to evaluate {TF-DCon}, mainly aiming to answer the following research questions: \textbf{RQ1}: how well the performance can be achieved when trained on datasets condensed by {TF-Con}? \textbf{RQ2}: how generalizable is the condensed datasets across different recommendation models? \textbf{RQ3}: How efficient is {TF-DCon} when compared to previous condensation methods? \textbf{RQ4}: {Can TF-DCon work well with other less powerful LLM?}
\subsection{Experimental Settings}
\subsubsection{Datasets} 
To evaluate the performance of our method, we condense the training set of three real-world datasets: MIND~\citep{mind2020acl}, Goodreads~\citep{goodreads2018Recsys}, and MovieLens~\citep{movieLens2015TITS}. The statistics of these datasets are shown in Table~\ref{tab:dataset-statistics}. We split the datasets into $80\%/10\%/10\%$ for training/validation/test, respectively. The condensation is mainly conducted on the training sets of datasets and during the testing phase, we still use the users and contents of items from the original datasets.
\subsubsection{Baselines} Due to the limited studies on condensation for CBR, we compare our proposed methods with three baselines implemented by ourselves: (i) \textit{Random}. In the setting of overall performance comparison (Table~\ref{tab:ablation-both}), we randomly select a certain portion of users and randomly select a certain portion of tokens to be the contents for each item. In the setting of user-level condensation (Table~\ref{tab:ablation-user}), it is implemented via randomly selecting certain ratios of users while the contents of items remain the same as those in original datasets. In the setting of content-level condensation (Table~\ref{tab:ablation-content}), we implement it by randomly selecting tokens for each item. (ii) \textit{Majority}. In the overall performance comparison, we implement this method by sampling the users with a large number of interactions and compressing the contents of items by randomly sampling the tokens. (iii) \textit{TextEmbed}. For the efficiency comparison, we adapt this text-embedding-based condensation method~\citep{lyq2021sigir-text-dd}, originally designed for classification, by replacing the cross-entropy loss with BPR loss. Due to its limited performance in our context, TextEmbed is included solely in the efficiency analysis and excluded in the main performance comparison. For fair comparison, we maintained an equivalent number of synthesized users as employed in our method, aiming to learn a representation for each user as well as a representation for the corresponding candidate item. 
\subsubsection{Evaluation Protocol}
To evaluate the quality of our method, we test the performance of recommendation models trained on the condensed datasets. In this evaluation, we employ three prevalent content-based recommendation models, i.e., NAML~\citep{ijcai19naml}, NRMS~\citep{nrms2019emnlp} and 
Fastformer~\citep{wu2021fastformer}. They are all designed in a hierarchical manner, i.e., using a content encoder to capture item representations in the history sequence, and a user encoder to fuse the historical item representations into a unified user representation.
To be specific, the evaluation is threefold: (1) condense the datasets, (2) train the recommendation models on the condensed datasets, and (3) test the performance of the models. For the evaluation metrics, we follow the common practice~\citep{nrms2019emnlp,ijcai19naml} to adopt the widely used metrics, i.e., NDCG@$K$ and Recall@$K$, abbreviated as N@$K$ and R@$K$ in the tables, respectively. In this work, we set $K = 1,5$ for evaluation on Goodreads and MovieLens, and set $K=5,10$ for MIND. To compare the performance of models trained on condensed datasets and original datasets, we average the percentages of metrics relative to the metrics trained on original datasets. We denote the percentages by ``Quality''.

\subsubsection{Implementation Details} 
During dataset condensation, we utilize GPT-3.5 for condensing. During training, we employ Adam~\citep{iclr2015adam} operator with a learning rate of 5e-3 for three datasets. We select title, abstract, and category feature of each piece of news as the item content for the MIND dataset, and select book or movie name and description feature of each book or movie as the item content for the Goodreads and MovieLens dataset. We concatenate multiple item contents into a single sequence before feeding them to the recommenders. For all three backbone models, we set the embedding dimension of content encoders to 256, the embedding dimension of user encoders to 64, and the negative sampling ratio to 4. We tune the hyperparameters of all base models to attain optimal performance. We average the results of five independent runs for each model. All experiments are conducted on a single NVIDIA GeForce RTX 3090 device.



\subsection{Overall Comparison (\textbf{RQ1})}
\newcommand{\first}[1]{{#1}}

\begin{table*}[]
\centering
\caption{{Overall Performance Comparison on Condensed Datasets.} ``Quality'' denotes the achieved ratio of performance when compared to those trained on original datasets. The detailed condensation ratio could be found in Table~\ref{tab:conden-ratio-overall}.}
\scalebox{0.705}{
\begin{threeparttable}          
\begin{tabular}{ll|llll|llll|llll}
\toprule[1.2pt]
\multicolumn{2}{l|}{{Datasets}}                                          & \multicolumn{4}{c|}{{MIND\tnote{1},}  $r$=5\%}                                                                                              & \multicolumn{4}{c|}{{Goodreads,} $r$=27\%}                                                                                         & \multicolumn{4}{c}{{MovieLens,} $r$=26\%}                                                                                          \\ \midrule
\multicolumn{1}{l|}{{Rec Model}}                    & {Metrics}   & {Random} & {Majority} & \cellcolor[HTML]{EFEFEF}{TF-DCon} & \textit{{Original}} & {Random} & {Majority} & \cellcolor[HTML]{EFEFEF}{TF-DCon} & \textit{{Original}} & {Random} & {Majority} & \cellcolor[HTML]{EFEFEF}{TF-DCon} & \textit{{Original}} \\ \midrule
\multicolumn{1}{l|}{}                                      & {N@1}       & 0.2871          & 0.2854            & \cellcolor[HTML]{EFEFEF}0.3071                               & \textit{0.3176}            & 0.5197          & 0.5057            & \cellcolor[HTML]{EFEFEF}0.5411                               & \textit{0.6462}            & 0.8241          & 0.8367            & \cellcolor[HTML]{EFEFEF}0.8484                               & \textit{0.828}             \\
\multicolumn{1}{l|}{}                                      & {N@5}       & 0.3470          & 0.3466            & \cellcolor[HTML]{EFEFEF}0.3691                               & \textit{0.3783}            & 0.7943          & 0.7884            & \cellcolor[HTML]{EFEFEF}0.8033                               & \textit{0.8475}            & 0.8251          & 0.8397            & \cellcolor[HTML]{EFEFEF}0.8494                               & \textit{0.831}             \\
\multicolumn{1}{l|}{}                                      & {R@1}       & 0.4016          & 0.4002            & \cellcolor[HTML]{EFEFEF}0.4377                               & \textit{0.4534}            & 0.4520          & 0.4326            & \cellcolor[HTML]{EFEFEF}0.4704                               & \textit{0.5635}            & 0.1785          & 0.1886            & \cellcolor[HTML]{EFEFEF}0.1873                               & \textit{0.1752}            \\
\multicolumn{1}{l|}{}                                      & {R@5}       & 0.5670          & 0.5697            & \cellcolor[HTML]{EFEFEF}0.6150                               & \textit{0.6270}            & 0.9983          & 0.9986            & \cellcolor[HTML]{EFEFEF}0.9984                               & \textit{0.9989}            & 0.7274          & 0.7323            & \cellcolor[HTML]{EFEFEF}0.7475                               & \textit{0.7399}            \\
\multicolumn{1}{l|}{\multirow{-5}{*}{{NAML}}}       & {Quality} & 90.28\%         & 90.15\%           & \cellcolor[HTML]{EFEFEF}\textbf{97.22\%}                     & \textit{100.00\%}          & 88.57\%         & 87.01\%           & \cellcolor[HTML]{EFEFEF}\textbf{90.49\%}                     & \textit{100.00\%}          & 99.75\%         & 102.18\%          & \cellcolor[HTML]{EFEFEF}\textbf{103.15\%}                    & \textit{100.00\%}          \\ \midrule
\multicolumn{1}{l|}{}                                      & {N@1}       & 0.2625          & 0.2631            & \cellcolor[HTML]{EFEFEF}0.2997                               & \textit{0.3009}            & 0.5399          & 0.5094            & \cellcolor[HTML]{EFEFEF}0.5453                               & \textit{0.6439}            & 0.8105          & 0.8294            & \cellcolor[HTML]{EFEFEF}0.8149                               & \textit{0.8178}            \\
\multicolumn{1}{l|}{}                                      & {N@5}       & 0.3225          & 0.3225            & \cellcolor[HTML]{EFEFEF}0.3597                               & \textit{0.3608}            & 0.8017          & 0.7901            & \cellcolor[HTML]{EFEFEF}0.8054                               & \textit{0.8476}            & 0.8227          & 0.8334            & \cellcolor[HTML]{EFEFEF}0.8371                               & \textit{0.8253}            \\
\multicolumn{1}{l|}{}                                      & {R@1}       & 0.3750          & 0.3793            & \cellcolor[HTML]{EFEFEF}0.4279                               & \textit{0.4325}            & 0.4704          & 0.4344            & \cellcolor[HTML]{EFEFEF}0.4751                               & \textit{0.5629}            & 0.1729          & 0.1821            & \cellcolor[HTML]{EFEFEF}0.1798                               & \textit{0.1757}            \\
\multicolumn{1}{l|}{}                                      & {R@5}       & 0.5414          & 0.5445            & \cellcolor[HTML]{EFEFEF}0.6000                               & \textit{0.6042}            & 0.9982          & 0.9990            & \cellcolor[HTML]{EFEFEF}0.9985                               & \textit{0.9993}            & 0.7325          & 0.7339            & \cellcolor[HTML]{EFEFEF}0.7446                               & \textit{0.7321}            \\
\multicolumn{1}{l|}{\multirow{-5}{*}{{NRMS}}}       & {Quality} & 88.23\%         & 88.66\%           & \cellcolor[HTML]{EFEFEF}\textbf{99.38\%}                     & \textit{100.00\%}          & 90.47\%         & 87.37\%           & \cellcolor[HTML]{EFEFEF}\textbf{91.01\%}                     & \textit{100.00\%}          & 99.31\%         & \textbf{101.57\%} & \cellcolor[HTML]{EFEFEF}101.28\%                             & \textit{100.00\%}          \\ \midrule
\multicolumn{1}{l|}{}                                      & {N@1}       & 0.2815          & 0.2736            & \cellcolor[HTML]{EFEFEF}0.3022                               & \textit{0.3057}            & 0.5420          & 0.5165            & \cellcolor[HTML]{EFEFEF}0.5548                               & \textit{0.6556}            & 0.7886          & 0.8105            & \cellcolor[HTML]{EFEFEF}0.8251                               & \textit{0.7915}            \\
\multicolumn{1}{l|}{}                                      & {N@5}       & 0.3425          & 0.3350            & \cellcolor[HTML]{EFEFEF}0.3637                               & \textit{0.3645}            & 0.8028          & 0.7934            & \cellcolor[HTML]{EFEFEF}0.8092                               & \textit{0.8529}            & 0.8254          & 0.8318            & \cellcolor[HTML]{EFEFEF}0.8429                               & \textit{0.8145}            \\
\multicolumn{1}{l|}{}                                      & {R@1}       & 0.3944          & 0.3804            & \cellcolor[HTML]{EFEFEF}0.4334                               & \textit{0.4365}            & 0.4725          & 0.4414            & \cellcolor[HTML]{EFEFEF}0.4851                               & \textit{0.5745}            & 0.1738          & 0.1799            & \cellcolor[HTML]{EFEFEF}0.1836                               & \textit{0.166}             \\
\multicolumn{1}{l|}{}                                      & {R@5}       & 0.5631          & 0.5515            & \cellcolor[HTML]{EFEFEF}0.6096                               & \textit{0.6144}            & 0.9983          & 0.9991            & \cellcolor[HTML]{EFEFEF}0.9986                               & \textit{0.9990}            & 0.739           & 0.7373            & \cellcolor[HTML]{EFEFEF}0.7465                               & \textit{0.7279}            \\
\multicolumn{1}{l|}{\multirow{-5}{*}{{Fastformer}}} & {Quality} & 92.01\%         & 89.58\%           & \cellcolor[HTML]{EFEFEF}\textbf{99.29\%}                     & \textit{100.00\%}          & 89.74\%         & 87.16\%           & \cellcolor[HTML]{EFEFEF}\textbf{90.97\%}                     & \textit{100.00\%}          & 101.80\%        & 103.55\%          & \cellcolor[HTML]{EFEFEF}\textbf{105.22\%}                    & \textit{100.00\%}          \\ \bottomrule[1.2pt]
\end{tabular}
\begin{tablenotes}
    \footnotesize
    \item[1] For the dataset MIND, we report $N@5, N@10$ and $R@5, R@10$, which is also the same in the subsequent tables of ablation studies.
\end{tablenotes}
\end{threeparttable}
}

\label{tab:ablation-both}
\end{table*}

\begin{table*}[]
\centering
\setlength\tabcolsep{2pt}

\caption{Comparison between condensed/sampled datasets and original datasets. We use ``ORI.'', ``RD.'' and ``MJ.'' to represent the statistics of original, random sampled, and majority-selected dataset. ``Item'' and ``User'' denote the condensation information of item contents and users, respectively.}
\label{tab:conden-ratio-overall}

\scalebox{0.9}{
\begin{tabular}{l|l|llll|llll|llll}
\toprule[1.2pt]
 \multicolumn{2}{l|}{{Datasets}}  & \multicolumn{4}{c|}{{MIND}} & \multicolumn{4}{c|}{{Goodreads}} & \multicolumn{4}{c}{{MovieLens}} \\
\midrule
\multicolumn{2}{l|}{{Methods}} & {OR.} & {RD.} & {MJ.} & {Ours} & {OR.} & {RD.} & {MJ.} & {Ours} & {OR.} & {RD.} & {MJ.} & {Ours} \\
\midrule
\multirow{3}{*}{{Item}} & {avg. tok.} 
& 45.42 & 16.73 & 16.73 & 16.73 
& 35.37 & 16.01 & 16.73 & 16.01 
& 34.80 & 15.26 & 16.73 & 15.26 \\
 & {Size (KB)} 
& 10,384 & 4,095 & 4,095 & 4,095 
& 2,189 & 993 & 993 & 993 
& 232 & 121 & 121 & 121 \\
 & {Ratio} 
& 100\% & 39\% & 39\% & 39\% 
& 100\% & 45\% & 45\% & 45\% 
& 100\% & 52\% & 52\% & 52\% \\
\midrule
\multirow{3}{*}{{User}} & {\#users} 
& 94,057 & 9,405 & 9,405 & 9,405 
& 23,089 & 4,617 & 2,350 & 4,617 
& 943 & 47 & 47 & 47 \\
 & {Size (KB)} 
& 106,614 & 2,124 & 5,733 & 2,139 
& 8,704 & 1,791 & 1,977 & 1,957
& 532 & 68 & 232 & 81 \\
 & {Ratio} 
& 100\% & 2\% & 5\% & 2\% 
& 100\% & 21\% & 23\% & 22\% 
& 100\% & 13\% & 44\% & 15\% \\
\midrule
\multirow{2}{*}{{Overall}} & {Size (KB)} 
& 116,998 & 6,219 & 9,828 & 6,234 
& 10,893 & 2,784 & 2,970 & 2,950 
& 764 & 189 & 353 & 202 \\
 & {Ratio} 
& 100\% & 5\% & 8\% & 5\% 
& 100\% & 26\% & 27\% & 27\% 
& 100\% & 25\% & 46\% & 26\% \\
\bottomrule[1.2pt]
\end{tabular}
}
\end{table*}
To validate the effectiveness of \ournameAbbr, we measure the recommendation performance of CBR models trained in the condensed datasets and their training efficiency. Specifically, we train the recommendation models (i.e., NAML, NRMS, Fastformer) on the condensed datasets and evaluate their performance on the test set of the original datasets. The results are reported in Table~\ref{tab:ablation-both} and the detailed condensation ratios are shown in Table~\ref{tab:conden-ratio-overall}. Further, we also evaluate the training efficiency on the condensed datasets and the results are summarized in Figure~\ref{fig:training-curve-on-datasets}.
\subsubsection{Overall Performance Comparison}  The proposed TF-DCon achieves better performance than the two baselines on three datasets. {We utilize ``'Quality'' to evaluate the quality of condensed dataset, which is computed by dividing each of the four metrics by the original performance, and then averaging them.} Notably, as shown in Table~\ref{tab:ablation-both}, we approximate over $97\%$ of the original performance with only $5\%$ data on the MIND dataset across three different recommender models. From Table~\ref{tab:conden-ratio-overall}, we observe that the amazing condensation ratio comes from two parts: reducing the size of contents by $61\%$ (item side) and reducing the size of historical interactions and users (user side)  by $98\%$. We observe similar performance on datasets Goodreads and MovieLens. Interestingly, when trained on condensed MovieLens, we may achieve a slightly better performance than those trained on the original dataset. We attribute this to the small size of the original MovieLens dataset since the condensed datasets may serve as a refined and denoised alternative to the original one. 

From Table~\ref{tab:conden-ratio-overall}, we can find that in each dataset, the size of the user part (i.e., users' historical interactions) is much larger than the size of the item part (contents of items). Consequently, the predominant size reduction stems from the condensation of the user part. 
We attribute this promising reduction to the common interests among different users, thus rendering that method of generating representative fake users to replace a variety users reasonably and effective. 
The condensation ratio of user-side on Goodreads and MovieLens dataset is less promising than the ratio on MIND. This discrepancy is ascribed to the fact that smaller datasets inherently exhibit fewer redundant interests among distinct users, consequently leading to a less pronounced reduction ratio. Specifically, within the MIND dataset, $M$ users may share analogous preferences and interaction histories. In contrast, this number diminishes to $N$ within the Goodreads and MovieLens datasets ($M>>N$). Consequently, in the condensed dataset derived from MIND, the synthesis of a single user effectively encapsulates the information equivalent to that of $M$ users in the original dataset. In comparison, the synthesis of one user in the condensed datasets derived from Goodreads and MovieLens covers the information of $N$ users in each dataset, respectively.

\subsubsection{Training Efficiency}
\begin{figure}
    \centering
    \setlength\tabcolsep{2pt}
    \begin{tabular}{ccc}
    \multicolumn{3}{c}{
        \resizebox{0.7\linewidth}{!}{
            \begin{tikzpicture}
    \begin{customlegend}[
        legend columns=3,
        legend style={
            align=left,
            draw=none,
            column sep=2ex
        },
        legend entries={
            \textsc{Original$_\text{NAML}$},
            \textsc{Original$_\text{NRMS}$},
            \textsc{Original$_\text{Fastformer}$},
            \textsc{TF-DCon$_\text{NAML}$},
            \textsc{TF-DCon$_\text{NRMS}$},
            \textsc{TF-DCon$_\text{Fastformer}$},
        }]
        \addlegendimage{olive,mark=x,dotted,line legend}
        \addlegendimage{orange,mark=x,dotted,line legend}
        \addlegendimage{purple,mark=x,dotted,line legend}
        \addlegendimage{olive,mark=x,solid,line legend}
        \addlegendimage{orange,mark=x,solid,line legend}
        \addlegendimage{purple,mark=x,solid,line legend}
        \end{customlegend}
\end{tikzpicture}
        }
    } \\
    \begin{subfigure}{0.3\linewidth}
        \resizebox{1.0\linewidth}{!}{
            \begin{tikzpicture}
\begin{axis}[    
    xlabel={Training Time (min)},
    ylabel={NDCG@5},    
    ymin=0.26,    
    ymax=0.32,
    xmax=42,
    grid=both,
    grid style=dashed,
]

    \addplot[olive,line width=2pt,dotted,mark=x] coordinates {

        (7.75, 0.2923)
        (15.50, 0.3074)
        (23.16, 0.3176)
        (30.84, 0.3113)
        (38.50, 0.3063)
    };

    \addplot[olive,line width=2pt,mark=x] coordinates {

        (1.86, 0.2947)
        (3.55, 0.2931)
        (5.16, 0.2992)
        (6.63, 0.3017)
        (7.95, 0.3079)
        (9.62, 0.3019)
    };

    \addplot[orange,line width=2pt,dotted,mark=x] coordinates {

        (18.66, 0.2970)
        (35.25, 0.3009)
        (50.55, 0.2851)
    };

    \addplot[orange,line width=2pt,mark=x] coordinates {

        (3.62, 0.2541)
        (7.00, 0.2972)
        (10.80, 0.2984)
        (13.62, 0.2997)
        (17.56, 0.2951)
    };

    \addplot[purple,line width=2pt,dotted,mark=x] coordinates {

        (17, 0.2825)
        (41, 0.3057)
        (63, 0.2895)
    };

    \addplot[purple,line width=2pt,mark=x] coordinates {

        (2.46, 0.2912)
        (4.63, 0.3022)
        (6.83, 0.3004)
    };
\end{axis}
\end{tikzpicture}
        }
        \caption{\label{fig:trend-mind}MIND}
    \end{subfigure} &
    \begin{subfigure}{0.315\linewidth}
        \resizebox{1.0\linewidth}{!}{
            \begin{tikzpicture}
\begin{axis}[    
    xlabel={Training Time (min)},
    ylabel={NDCG@5},    
    ymin=0.74,    
    ymax=0.86,
    xmax=100,
    grid=both,
    grid style=dashed,
]

    \addplot[olive,line width=2pt,dotted,mark=x] coordinates {
        (5.67, 0.8066)
        (11.50, 0.8151)
        (17.17, 0.8199)
        (21.66, 0.8231)
        (27.17, 0.8269)
        (31.60, 0.8280)
        (36.00, 0.8301)
        (40.50, 0.8322)
        (44.00, 0.8338)
        (48.83, 0.8352)
        (53.60, 0.8351)
        (58.38, 0.8382)
        (62.75, 0.8381)
        (66.05, 0.8380)
        (70.30, 0.8412)
    };
    \addplot[olive,line width=2pt,mark=x] coordinates {
        (1.07, 0.7831)
        (1.97, 0.7921)
        (2.87, 0.7940)
        (3.83, 0.7982)
        (4.78, 0.7982)
        (5.63, 0.7982)
        (6.38, 0.7999)
        (7.15, 0.8001)
        (7.87, 0.8005)
        (8.55, 0.8019)
        (9.17, 0.8019)
        (9.83, 0.8019)
        (10.78, 0.8028)
        (11.72, 0.8018)
        (12.55, 0.8024)
        (13.38, 0.8033)
        (14.20, 0.8027)
        (15.03, 0.8033)
        (15.88, 0.8028)
    };

    \addplot[orange,line width=2pt,dotted,mark=x] coordinates {
        (6.60, 0.8151)
        (13.07, 0.8226)
        (19.65, 0.8277)
        (26.17, 0.8337)
        (32.72, 0.8356)
        (39.30, 0.8396)
        (45.87, 0.8442)
        (52.45, 0.8470)
        (59.07, 0.8457)
        (65.65, 0.8487)
    };

    \addplot[orange,line width=2pt,mark=x] coordinates {
        (0.87, 0.7711)
        (1.55, 0.7853)
        (2.42, 0.7883)
        (3.20, 0.7919)
        (3.92, 0.7921)
        (4.75, 0.7950)
        (5.53, 0.7975) 
        (6.28, 0.7982)
        (7.12, 0.8001)
        (8.63, 0.8010)
        (9.47, 0.8022)
        (10.18, 0.8020)
        (11.02, 0.8033)
        (11.82, 0.8040)
        (12.53, 0.8039)
        (13.37, 0.8025)
        (14.17, 0.8042)
        (14.98, 0.8053)
        (15.76, 0.8051)
        (16.56, 0.8044)
    };
     
    \addplot[purple,line width=2pt,dotted,mark=x] coordinates {
        (8.02, 0.8105)
        (15.93, 0.8256)
        (23.87, 0.8310)
        (31.82, 0.8330)
        (39.68, 0.8367)
        (47.57, 0.8441)
        (55.38, 0.8446)
        (63.23, 0.8452)
        (71.08, 0.8493)
        (78.97, 0.8484)
        (86.85, 0.8515)
        (94.72, 0.8539)
        (102.62, 0.8542)
        (110.48, 0.8565)
        (118.32, 0.8548)
    };

    \addplot[purple,line width=2pt,mark=x] coordinates {
        (1.00, 0.7660)
        (1.98, 0.7737)
        (3.00, 0.7812)
        (4.02, 0.7880)
        (5.02, 0.7933)
        (5.93, 0.7919)
        (6.90, 0.7945)
        (7.92, 0.7943)
        (8.93, 0.7970)
        (9.92, 0.7948)
        (10.82, 0.7988)
        (11.80, 0.8013)
        (12.83, 0.8024)
        (13.83, 0.8033)
        (14.78, 0.8030)
        (15.65, 0.8046)
        (16.62, 0.8054)
        (17.63, 0.8063)
        (18.65, 0.8055)
        (19.60, 0.8060)
    };
    
\end{axis}
\end{tikzpicture}
        }
        \caption{\label{fig:trend-goodreads}Goodreads}
    \end{subfigure}
    &
    \begin{subfigure}{0.315\linewidth}
        \resizebox{0.98\linewidth}{!}{
            \begin{tikzpicture}
\begin{axis}[    
    xlabel={Training Time (min)},
    ylabel={NDCG@5},    
    ymin=0.81,    
    ymax=0.87,
    xmax=4,
    grid=both,
    grid style=dashed,
]

    \addplot[olive,line width=2pt,dotted,mark=x] coordinates {
        (0, 0)
        (0.53, 0.8422)
        (1.22, 0.8544)
        (1.92, 0.8533)
    };

    \addplot[olive,line width=2pt,mark=x] coordinates {
        (0, 0)
        (0.37, 0.8531)
        (0.73, 0.8563)
        (1.1, 0.8574)
    };

    \addplot[orange,line width=2pt,dotted,mark=x] coordinates {
        (0, 0)
        (1.73, 0.8433)
        (2.67, 0.8458)
        (3.93, 0.8429)
    };

    \addplot[orange,line width=2pt,mark=x] coordinates {
        (0, 0)
        (0.82, 0.8491)
        (1.65, 0.856)
        (2.47, 0.8513)
    };

    \addplot[purple,line width=2pt,dotted,mark=x] coordinates {
        (0, 0)
        (1.99, 0.8309)
        (3.58, 0.8416)
        (5.22, 0.8321)
    };

    \addplot[purple,line width=2pt,mark=x] coordinates {
        (0, 0)
        (1.45, 0.8323)
        (3.47, 0.8382)
        (5.2, 0.83)
    };
\end{axis}
\end{tikzpicture}
        }
        \caption{\label{fig:trend-movielen}{MovieLens}}
    \end{subfigure}
    \end{tabular}
    \caption{{Training efficiency on original and condensed datasets.}}
    
    \label{fig:training-curve-on-datasets}
    \vskip -0.1in
\end{figure}
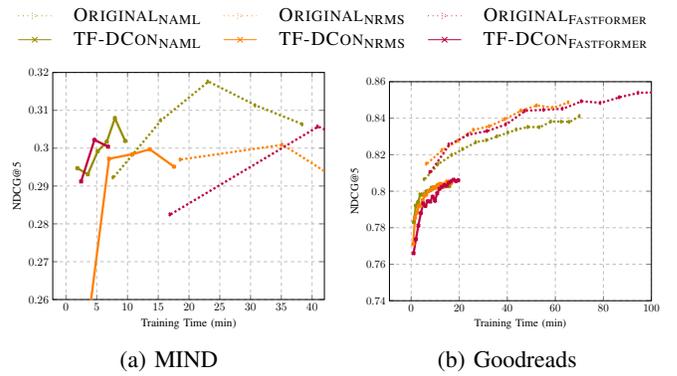
Here, we compare the training time and performance of NAML, NRMS and Fastformer when trained on the original datasets and condensed datasets by TF-DCon. As shown in Figure~\ref{fig:training-curve-on-datasets}, the solid lines represent the training curves on condensed datasets and the dotted line represents original datasets, where we can observe that the training on condensed datasets converges much faster than the training on original datasets (i.e., up to $5\times$ speedup). Typically, the model convergences on condensed datasets are reached with around $6$ minutes while the training on original datasets needs around or more than $30$ minutes. For the performance, we find that those trained on condensed datasets can exhibit comparability with regard to the metric NDCG@5.
\subsection{{Generalization Across Recommender Models (RQ2)}}
\subsubsection{Generalizability of Condensed Datasets} We test the performance of CBR models with different architectures trained on the datasets condensed by one CBR model. Specifically, we employ NAML~\citep{ijcai19naml} as the backbone model for condensation and test the performance of NAML~\citep{ijcai19naml}, NRMS~\citep{nrms2019emnlp} and Fastformer~\citep{wu2021fastformer} trained on the condensed datasets. The results are reported in Table~\ref{tab:ablation-both}. Furthermore, we also test the generalizability of content-level and user-level condensation in the ablation study for \ournameAbbr, as shown in Table~\ref{tab:ablation-content} and Table~\ref{tab:ablation-user}. From the results, we can observe that the datasets condensed by \ournameAbbr exhibit good generalization on different CBR models. We ascribe this capability to textual data, as texts exclusively encapsulate semantic information while excluding architectural details found in text embeddings.
\subsubsection{Versatility of \ournameAbbr} The proposed \ournameAbbr is highly flexible since we can adopt different CBR models as the user encoder in the condensation method. We investigate the performances of different models on datasets condensed by \ournameAbbr with different recommendation models. We utilize NAML, NRMS, and Fastformer as the user encoder for both condensation and evaluation, respectively. We choose the MIND dataset to report results in Table~\ref{tab:generalization-tf-con}, where we can observe the strong generalization ability of condensed datasets on other models. We can also observe that NAML can outperform the other two models when trained on the most condensed dataset as well as the original dataset.
\begin{table}[]
\centering
\caption{\ournameAbbr with different recommender models for condensation. The datasets condensed by different recommender models exhibit similar training performance across different recommender models, verifying the versatility of our method.}

\begin{tabular}{l|ll|ll|ll|ll}
\toprule
{Train Set} & \multicolumn{2}{l|}{{\ournameAbbr$_{NAML}$}} & \multicolumn{2}{l|}{{\ournameAbbr$_{NRMS}$}} & \multicolumn{2}{l|}{{\ournameAbbr$_{Fast.}$}} & \multicolumn{2}{l}{{Original}} \\
\midrule
{Metrics} & {N@5} & {R@5} & {N@5} & {R@5} & {N@5} & {R@5} & {N@5} & {R@5} \\
\midrule
{NAML} & 0.3071 & 0.4377 & 0.3066 & 0.4396 & 0.3060 & 0.4368 & {0.3176} & {0.4534} \\
{NRMS} & 0.2997 & 0.4279 & 0.2905 & 0.4256 & 0.2924 & 0.4184 & {0.3009} & {0.4325} \\
{Fastformer} & 0.3022 & 0.4334 & {0.3094} & {0.4409} & 0.3056 & 0.4343 & 0.3057 & 0.4365 \\
\bottomrule
\end{tabular}
\label{tab:generalization-tf-con}
\end{table}

\subsection{Condensation Efficiency (RQ3)}
Since TF-DCon is a training-free condensation method, we examine its efficiency and compare it with previous condensation method in this section. Specifically, we compare the performance and running time of TF-DCon with TextEmbed~\citep{lyq2021sigir-text-dd}, which follows the most prevalent optimization paradigm~\citep{zhaoICLR2021DC,jinICLR2022graphCondesation}.
We report the results on three datasets with NAML as the recommendation model, as shown in Table~\ref{tab:compare-with-textEmbedd}. We can observe that TF-DCon is much faster than the iterative TextEmbed and the performance is also more superior. The efficiency of TF-DCon originates from its training-free design, which is free from the nested optimization involved in TextEmbed.
\subsection{{Generalization Across Language Models(RQ4)}}
In this section, we aim to examine: 
\begin{itemize}
    \item The \textbf{generalization of TF-DCon with other language models}.
    \item The \textbf{impact of evolutionary prompt generations} on condensation performance.
\end{itemize}
Specifically, we equip TF-DCon with Llama-2-7b~\citep{Llama-2} and utilize different generation of prompts from EvoPro (Algorithm~\ref{alg:EvoPro}) for the content-level condensation. 

To examine the performance of TF-DCon with Llama-2 and the influence of EvoPro, we conduct experiments on three datasets with recommendation model NAML. The results are reported in Table~\ref{tab:openllm-evo-prompt}, where ``-evo-0'' denotes the method with the initial handcrafted prompt and ``-evo-i'' denotes the method with \textit{$gen_{i}$-prompt}. On the one hand, we can observe that the method ``TF-DCon'' with ChatGPT outperforms all the variants with Llama-2, which is reasonable since the capability of text comprehension and generation of ChatGPT is superior to Llama-2-7b. However, the performance with Llama-2-7b (``-evo-i'') remains close to that achieved with ChatGPT (``TF-DCon''), indicating that our method is effective even with smaller-scale language models. On the other hand, we can observe that the next generation of prompt can consistently outperform its ``parent'': ``-evo-1'' is better than ``-evo-0'' while ``-evo-2'' is better than ``-evo-1''. This demonstrates that the evolution of the prompts is beneficial in adapting LLM for condensation.
\begin{table}[]
\caption{{TF-DCon with Open-sourced LLM and Evolutionary Prompting for Condensation.}}
\scalebox{0.93}{\begin{tabular}{l|llll|llll|llll}
\toprule
Dataset & \multicolumn{4}{c|}{MIND, $r$=5\%}             & \multicolumn{4}{c|}{Goodreads, $r$=27\%} & \multicolumn{4}{c}{MovieLens, $r$=26\%}         \\ \midrule
Metric  & N@1 & N@5 & R@1 & R@5 & N@1           & N@5           & R@1           & R@5           & N@1 & N@5 & R@1 & R@5 \\ \midrule
-evo-0 &
  0.2902 &
  0.3494 &
  0.4278 &
  0.5919 &
   0.5098 &
   0.7903 &
   0.4419 &
   0.9978 &
  0.8330 &
  0.8359 &
  0.1802 &
  0.7312 \\
-evo-1  & 0.2934 & 0.3526 & 0.4315   & 0.6016   & 0.5122           & 0.7951           & 0.4546             & 0.9979             & 0.8378 & 0.8410  & 0.1862   & 0.7382   \\
-evo-2  & 0.2993 & 0.3545 & 0.4329   & 0.6021   & 0.5258           & 0.7981           & 0.4642             & 0.9982             & 0.8423 & 0.8464 & 0.1857   & 0.7416   \\
TF-DCon & 0.3071 & 0.3691 & 0.4377   & 0.6150    & 0.5411           & 0.8033           & 0.4704             & 0.9984             & 0.8480  & 0.8494 & 0.1873   & 0.7475   \\ \bottomrule
\end{tabular}}
\label{tab:openllm-evo-prompt}
\end{table}
\begin{table}[]
\caption{{Efficiency and performance comparison between embedding-based method and TF-DCon.}}
\scalebox{0.83}{\begin{tabular}{l|llll|llll|llll}
\toprule
Dataset &
  \multicolumn{4}{c|}{MIND} &
  \multicolumn{4}{c|}{Goodreads} &
  \multicolumn{4}{c}{MovieLens} \\ \midrule
Methods &
  \multicolumn{2}{l|}{TextEmbedd} &
  \multicolumn{2}{l|}{TF-DCon} &
  \multicolumn{2}{l|}{TextEmbedd} &
  \multicolumn{2}{l|}{TF-DCon} &
  \multicolumn{2}{l|}{TextEmbedd} &
  \multicolumn{2}{l}{TF-DCon} \\ \midrule
N@1 &
  \multicolumn{1}{l|}{0.2678} &
  \multicolumn{1}{l|}{\multirow{4}{*}{450.02min}} &
  \multicolumn{1}{l|}{0.3071} &
  \multirow{4}{*}{67.46min} &
  \multicolumn{1}{l|}{0.5013} &
  \multicolumn{1}{l|}{\multirow{4}{*}{126.24min}} &
  \multicolumn{1}{l|}{0.5411} &
  \multirow{4}{*}{6.15min} &
  \multicolumn{1}{l|}{0.7128} &
  \multicolumn{1}{l|}{\multirow{4}{*}{64.58min}} &
  \multicolumn{1}{l|}{0.8484} &
  \multirow{4}{*}{0.47min} \\
N@5 &
  \multicolumn{1}{l|}{0.3147} &
  \multicolumn{1}{l|}{} &
  \multicolumn{1}{l|}{0.3691} &
   &
  \multicolumn{1}{l|}{0.7794} &
  \multicolumn{1}{l|}{} &
  \multicolumn{1}{l|}{0.8033} &
   &
  \multicolumn{1}{l|}{0.7569} &
  \multicolumn{1}{l|}{} &
  \multicolumn{1}{l|}{0.8494} &
   \\
R@1 &
  \multicolumn{1}{l|}{0.3815} &
  \multicolumn{1}{l|}{} &
  \multicolumn{1}{l|}{0.4377} &
   &
  \multicolumn{1}{l|}{0.4301} &
  \multicolumn{1}{l|}{} &
  \multicolumn{1}{l|}{0.4704} &
   &
  \multicolumn{1}{l|}{0.1474} &
  \multicolumn{1}{l|}{} &
  \multicolumn{1}{l|}{0.1873} &
   \\
R@5 &
  \multicolumn{1}{l|}{0.5198} &
  \multicolumn{1}{l|}{} &
  \multicolumn{1}{l|}{0.6150} &
   &
  \multicolumn{1}{l|}{0.9953} &
  \multicolumn{1}{l|}{} &
  \multicolumn{1}{l|}{0.9984} &
   &
  \multicolumn{1}{l|}{0.6884} &
  \multicolumn{1}{l|}{} &
  \multicolumn{1}{l|}{0.7475} &
   \\ \bottomrule
\end{tabular}}
\vskip -0.05in
\label{tab:compare-with-textEmbedd}
\end{table}

\begin{table}[]
\caption{{Analysis on the number of cluster K.}}
\scalebox{0.9}{\begin{tabular}{l|llll|llll|llll}
\toprule
Dataset & \multicolumn{4}{c|}{MIND}         & \multicolumn{4}{c|}{Goodreads}    & \multicolumn{4}{c}{MovieLens}     \\ \midrule
K       & 1154   & 2308   & 4617   & 5000   & 1154   & 2308   & 4617   & 5000   & 1154   & 2308   & 4617   & 5000   \\ \midrule
N@1     & 0.2846 & 0.2913 & 0.3071 & 0.3088 & 0.5005 & 0.522  & 0.5411 & 0.5467 & 0.8196 & 0.8345 & 0.8484 & 0.8474 \\
N@5     & 0.3478 & 0.356  & 0.3691 & 0.3701 & 0.7864 & 0.7959 & 0.8033 & 0.8017 & 0.8213 & 0.8399 & 0.8494 & 0.8486 \\
R@1     & 0.4014 & 0.4135 & 0.4377 & 0.4397 & 0.4344 & 0.4541 & 0.4704 & 0.4804 & 0.1779 & 0.1801 & 0.1873 & 0.1893 \\
R@5     & 0.5686 & 0.5997 & 0.615  & 0.6204 & 0.9984 & 0.9983 & 0.9984 & 0.9986 & 0.7304 & 0.7336 & 0.7475 & 0.762  \\ \bottomrule
\end{tabular}}
\label{tab:cluster-k}
\vskip -0.1in
\end{table}
\subsection{Ablation Study} 
To examine how \textit{user-level condensation} and \textit{content-level condensation} affect the overall performance, we performed an ablation study on them, respectively. We examine the performance of models trained on datasets generated by only user-level condensation or content-level condensation. 

Specifically, content-level condensation only condense the contents of items and the historical interactions. The results are reported in Table~\ref{tab:ablation-content}, where we observe that only condensing the item contents can also achieve comparable performance. The condensation ratio is relatively not promising and we attribute this to the information capacity of the contents which restricts the size of textual data. For user-level condensation, we only condense the users into fake users and their corresponding interactions while the contents are preserved. As shown in Table~\ref{tab:ablation-user}, we can find that the condensation ratio is astonishingly good while we can still achieve good performance. Especially, on the MIND dataset, we can achieve over 97\% performance of the original datasets with 98\% size reduction of the user-side dataset (i.e., the historical interactions of users ) when tested on Fastformer. When tested on NAML and NRMS, the performance can also exceed over 93\% of those trained on the original dataset. Besides, the condensation ratios on Goodreads and MovieLens are also good with a comparable performance. We attribute the promising condensation ratios to similar interests among different users. Therefore, when we synthesize a fake user to represent a group of users with common interests, we can provide adequate information for model training.

{Moreover, by comparing the performance of different methods in content-level condensation (Table~\ref{tab:ablation-content}) to the performance in user-level condensation (Table~\ref{tab:ablation-user}) and both-level condensation (Table~\ref{tab:ablation-both}), we can find that TF-DCon consistently maintains superior performance while the baseline methods suffer a large performance drop while reducing the user number. We attribute this to the ability of TF-DCon in preserving the preference information for most users via the user synthesis.}
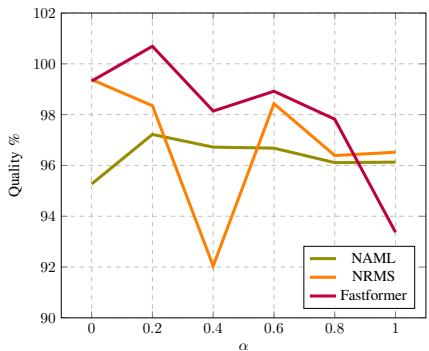
\begin{figure}
    \centering
    \setlength\tabcolsep{2pt}
    \begin{tabular}{ccc}
    \multicolumn{3}{c}{
        \resizebox{0.65\linewidth}{!}{
            \begin{tikzpicture}
    \begin{customlegend}[
        legend columns=3,
        legend style={
            align=left,
            draw=none,
            column sep=2ex
        },
        legend entries={
            \textsc{NAML},
            \textsc{NRMS},
            \textsc{Fastformer},
        }]
        \addlegendimage{olive,mark=x,solid,line legend}
        \addlegendimage{orange,mark=x,solid,line legend}
        \addlegendimage{purple,mark=x,solid,line legend}
        \end{customlegend}
\end{tikzpicture}
        }
    } \\
    \begin{subfigure}{0.315\linewidth}
        \resizebox{1.0\linewidth}{!}{
            \begin{tikzpicture}
\begin{axis}[    
    xlabel={$\alpha$},
    ylabel={Quality \%},    
    ymin=93.5,    
    ymax=101,     
    grid=both,
    grid style=dashed,
]

    \addplot[olive,line width=2pt] coordinates {
        (0,	95.378033)
        (0.2,	97.3315318)
        (0.4,	96.8530091)
        (0.6,	96.8248607)
        (0.8,	96.3407082)
        (1,	96.2844114)
    };

     \addplot[orange,line width=2pt] coordinates {
        (0,	94.6203486)
        (0.2,	98.6203486)
        (0.4,	99.3464437)
        (0.6,	97.8844114)
        (0.8,	96.8844114)
        (1,	94.8844114)
    };
    
    \addplot[purple,line width=2pt] coordinates {
        (0,	96.2749985)
        (0.2,	97.9749985)
        (0.4,	98.1749985)
        (0.6,	99.291151)
        (0.8,	95.6749985)
        (1,	94.6749985)
    };

\end{axis}
\end{tikzpicture}
        }
        \caption{\label{fig:alpha-mind}{MIND}}
    \end{subfigure} 
    &
    \begin{subfigure}{0.315\linewidth}
        \resizebox{1.0\linewidth}{!}{
            \begin{tikzpicture}
\begin{axis}[    
    xlabel={$\alpha$},
    ylabel={Quality \%},    
    ymin=85.5,    
    ymax=92,     
    grid=both,
    grid style=dashed,
]

    \addplot[olive,line width=2pt] coordinates {
        (0,	88.16271365)
        (0.2,	89.92111795)
        (0.4,	90.49)
        (0.6,	88.60828001)
        (0.8,	89.04901269)
        (1,	87.44859287)
    };

     \addplot[orange,line width=2pt] coordinates {
        (0,	87.26610594)
        (0.2,	89.18479303)
        (0.4,	91.01)
        (0.6,	90.27966804)
        (0.8,	88.61902707)
        (1,	87.00340444)
    };
    
    \addplot[purple,line width=2pt] coordinates {
        (0,	88.08783413)
        (0.2,	89.60095012)
        (0.4,	90.08114268)
        (0.6,	90.97)
        (0.8,	88.06357951)
        (1,	87.493437)
    };

\end{axis}
\end{tikzpicture}
        }
        \caption{\label{fig:alpha-goodreads}{Goodreads}}
    \end{subfigure}
    &
    \begin{subfigure}{0.315\linewidth}
        \resizebox{1\linewidth}{!}{
            \begin{tikzpicture}
\begin{axis}[    
    xlabel={$\alpha$},
    ylabel={Quality \%},    
    ymin=90,    
    ymax=102,     
    grid=both,
    grid style=dashed,
]

    \addplot[olive,line width=2pt] coordinates {
        (0.0, 95.27)
        (0.2, 97.22)
        (0.4, 96.72)
        (0.6, 96.68)
        (0.8, 96.11)
        (1.0, 96.13)
    };

     \addplot[orange,line width=2pt] coordinates {
        (0.0, 99.38)
        (0.2, 98.35)
        (0.4, 92.04)
        (0.6, 98.43)
        (0.8, 96.39)
        (1.0, 96.52)
    };
    
    \addplot[purple,line width=2pt] coordinates {
        (0.0, 99.32)
        (0.2, 100.69)
        (0.4, 98.14)
        (0.6, 98.92)
        (0.8, 97.82)
        (1.0, 93.37)
    };

\end{axis}
\end{tikzpicture}
        }
        \caption{\label{fig:alpha-movielen}{MovieLens}}
    \end{subfigure}
    \end{tabular}
    \caption{{Varying the hyperparameter $\alpha$ in Eq.~\ref{eq:dis-user}.}}
    
    \label{fig:alpha}
    \vskip -0.1in
\end{figure}
\begin{table}[]
\centering
\vskip -0.1in
\caption{Ablation Study on Content-Level Condensation.}

\scalebox{0.9}{
\begin{tabular}{l|l|lll|lll|lll}
\toprule[1.2pt]
 \multicolumn{2}{l|}{{Datasets}}  & \multicolumn{3}{c|}{{MIND,} $r=39\%$} & \multicolumn{3}{c|}{{Goodreads,} $r=45\%$} & \multicolumn{3}{c}{{MovieLens,} $r=52\%$} \\
\midrule
{Rec Model}& {Metrics} & {Random} & {\ournameAbbr} & {Original} & {Random} & {\ournameAbbr} & {Original} & {Random} & {\ournameAbbr} & {Original} \\
\midrule
\multirow{6}{*}{{NAML}} 
 & {N@1} 
 & 0.3059 & 0.3126 & 0.3176 
 & 0.5807 & 0.6359 & 0.6462 
 & 0.7959 & 0.8382 & 0.8280 \\
 & {N@5} 
 & 0.3678 & 0.3725 & 0.3783 
 & 0.8204 & 0.8434 & 0.8475 
 & 0.8208 & 0.8367 & 0.8310 \\
 & {R@1} 
 & 0.4396 & 0.4457 & 0.4534 
 & 0.5053 & 0.5543 & 0.5635 
 & 0.1665 & 0.1811 & 0.1752 \\
 & {R@5} 
 & 0.6167 & 0.6169 & 0.6270 
 & 0.9988 & 0.9989 & 0.9989 
 & 0.7341 & 0.7374 & 0.7399 \\
 & {Quality} 
 & 97.39\% & 98.39\% & 100\%  
 & 95.06\% & 99.23\% & 100\% 
 & 97.79\% & 100.75\% & 100\% \\ 
\midrule
\multirow{6}{*}{{NRMS}} 
& {N@1} 
 & 0.2902 & 0.3050 & 0.3009 
 & 0.6320 & 0.6668 & 0.6439
 & 0.8017 & 0.8207 & 0.8178 \\
 & {N@5} 
 & 0.3525 & 0.3626 & 0.3608
 & 0.8421 & 0.8561 & 0.8476
 & 0.8250 & 0.8285 & 0.8253 \\
 & {R@1} 
 & 0.4205 & 0.4387 & 0.4325
 & 0.5521 & 0.5832 & 0.5629
 & 0.1724 & 0.1759 & 0.1757 \\
 & {R@5} 
 & 0.5998 & 0.6041 & 0.6042
 & 0.9991 & 0.9990 & 0.9993
 & 0.7329 & 0.7354 & 0.7321 \\
 & {Quality} 
 & 97.23\% & 99.30\% & 100\% 
 & 97.43\% & 98.34\% & 100\% 
 & 99.26\% & 100.38\% & 100\% \\ 
\midrule
\multirow{6}{*}{{Fastformer}} 
 & {N@1} 
 & 0.2988 & 0.3025 & 0.3057 
 & 0.6296 & 0.6499 & 0.6556 
 & 0.7828 & 0.7974 & 0.7915 \\
 & {N@5} 
 & 0.3599 & 0.3643 & 0.3645 
 & 0.8408 & 0.8506 & 0.8529
 & 0.8086 & 0.8159 & 0.8145 \\
 & {R@1} & 0.4321 & 0.4359 & 0.4365 
 & 0.5495 & 0.5701 & 0.5745
 & 0.1653 & 0.1699 & 0.1660 \\
 & {R@5} & 0.6067 & 0.6129 & 0.6144 
 & 0.9991 & 0.9991 & 0.9990
 & 0.7223 & 0.7266 & 0.7279 \\
 & {Quality} 
 & 98.63\% & 99.67\% & 100\% 
 & 97.96\% & 99.60\% & 100\% 
 & 99.16\% & 100.40\% & 100\% \\ 
\bottomrule[1.2pt]
\end{tabular}}
\label{tab:ablation-content}
\end{table}

\begin{table}[]
\centering
\vskip -0.1in
\caption{Ablation Study on User-Level Condensation.}

\scalebox{0.9}{\begin{tabular}{l|l|lll|lll|lll}
\toprule[1.2pt]
 \multicolumn{2}{l|}{{Datasets}}  & \multicolumn{3}{c|}{{MIND,} $r=2\%$} & \multicolumn{3}{c|}{{Goodreads,} $r=22\%$} & \multicolumn{3}{c}{{MovieLens,} $r=15\%$} \\
\midrule
{Rec Model}& {Metrics} & {Random} & {\ournameAbbr} & {Original} & {Random} & {\ournameAbbr} & {Original} & {Random} & {\ournameAbbr} & {Original} \\
\midrule
\multirow{6}{*}{{NAML}} 
 & {N@1} 
    & 0.2886 & 0.2964 & 0.3176 
    & 0.5259 & 0.5391 & 0.6462
    & 0.8154 & 0.8338 & 0.8280 \\
 & {N@5} 
    & 0.3474 & 0.3528 & 0.3783 
    & 0.7970 & 0.8034 & 0.8475 
    & 0.8350 & 0.8470 & 0.8310 \\
 & {R@1} 
    & 0.3995 & 0.4273 & 0.4534 
    & 0.4570 & 0.4736 & 0.5635 
    & 0.1776 & 0.1879 & 0.1752 \\
 & {R@5} 
    & 0.5627 & 0.5887 & 0.6270 
    & 0.9983 & 0.9983 & 0.9989 
    & 0.7457 & 0.7468 & 0.7399 \\
 & {Quality}
    & 89.97\% & 93.75\% & 100\% 
    & 90.91\% & 92.09\% & 100\%
    & 99.98\% & 101.61\% & 100\% \\
\midrule
\multirow{6}{*}{{NRMS}} 
 & {N@1} 
    & 0.2702 & 0.2834 & 0.3009 
    & 0.5488 & 0.5559 & 0.6439 
    & 0.8132 & 0.8207 & 0.8178 \\
 & {N@5} 
    & 0.3281 & 0.3449 & 0.3608 
    & 0.8066 & 0.8099 & 0.8476 
    & 0.8208 & 0.8338 & 0.8253 \\
 & {R@1} 
    & 0.3796 & 0.4159 & 0.4325 
    & 0.4793 & 0.4851 & 0.5629 
    & 0.1738 & 0.1802 & 0.1757 \\
 & {R@5} 
    & 0.5413 & 0.5927 & 0.6042 
    & 0.9984 & 0.9984 & 0.9993
    & 0.7291 & 0.7373 & 0.7321 \\
 & {Quality}
    & 89.45\% & 96.38\% & 100\% 
    & 92.78\% & 93.31\% & 100\%
    & 99.45\% & 100.83\% & 100\% \\
\midrule
\multirow{6}{*}{{Fastformer}} 
 & {N@1} 
    & 0.2880 & 0.2978 & 0.3057 
    & 0.5564 & 0.5614 & 0.6556 
    & 0.8047 & 0.8251 & 0.7915 \\
 & {N@5} 
    & 0.3469 & 0.3568 & 0.3645 
    & 0.8094 & 0.8113 & 0.8529 
    & 0.8260 & 0.8412 & 0.8145 \\
 & {R@1} 
    & 0.3931 & 0.4255 & 0.4365 
    & 0.4860 & 0.4902 & 0.5745 
    & 0.1740 & 0.1815 & 0.1660 \\
 & {R@5} 
    & 0.5575 & 0.5952 & 0.6144 
    & 0.9985 & 0.9983 & 0.9990 
    & 0.7359 & 0.7460 & 0.7279 \\
 & {Quality}
    & 92.12\% & 97.34\% & 100\% 
    & 92.48\% & 92.84\% & 100\%
    & 101.63\% & 103.76\% & 100\% \\
\bottomrule[1.2pt]
\end{tabular}}
\label{tab:ablation-user}
\vskip -0.1in
\end{table}

\subsection{Further Investigation}
\subsubsection{Analysis on hyperparameter $\alpha$.} In this section, we further investigate how different values of hyperparameter $\alpha$ affect the quality of the condensed datasets. 
The hyperparameter $\alpha$ is utilized to balance the influence of user interests and user embeddings on the user synthesis. The larger value of $\alpha$ denotes larger influence of user interests on the synthesis of fake users. As shown in Figure~\ref{fig:alpha}, we can find that in most cases, the best choice for $\alpha$ with three different recommendation models is smaller than 1 but larger than 0, which verifies that the effectiveness of selecting/merging users with similar interests for the synthesis. {Besides, the selection of $\alpha$ varies among three recommendation models, influenced by their respective interest-capturing ability. A smaller optimal $\alpha$ indicates the model's diminished capacity to capture information from limited texts, such as user interests. We can observe that Fastformer possesses the best interest-capturing ability, which is also in line with the results in previous study~\cite{wu2021fastformer}.}
\subsubsection{Analysis on the value of K}
In this section, we study how the value of $K$ affect the condensation performance. To be noted, the $K$ equals the number of synthesized users in the condensed dataset. Therefore, increasing the number of $K$ will increase the size of the condensed dataset. Specifically, we conduct experiment on three datasets with the recommendation model NAML. The result is reported in Table~\ref{tab:cluster-k}, where we can observe that increasing $K$ can lead to better performance of NAML, denoting better quality of the condensed dataset.
\subsubsection{Effectiveness of User Interests} In this section, to examine the effectiveness of the user interests in assisting the user condensation, conduct an ablation study on the MovieLens dataset {by designing a variant of TF-DCon, denoted by ``$\text{TF-DCon}_{RI}$''. We implement ``$\text{TF-DCon}_{RI}$'' by} replacing the user interests with tokens randomly sampled from his/her historical items and restricting the number of the tokens to the same number of user interests. The results are shown in Table~\ref{tab:ablation-user-interest}. We can observe that randomly sampled tokens under-perform TF-DCon, which demonstrates the effectiveness of our design of ChatGPT-enhanced interests extraction module.
\begin{table}[]
\vskip -0.1in
\caption{{Ablation Study on User Interests.}}
\scalebox{0.9}{
\begin{tabular}{l|l|l|llll}
\toprule
Dataset                    & Rec Model                   & Metric          & N@1    & N@5    & R@1    & R@5    \\ \midrule
\multirow{6}{*}{MIND}      & \multirow{2}{*}{NAML}       & $\text{TF-DCon}_{RI}$ & 0.2978 & 0.3647 & 0.4203 & 0.5889 \\
                           &                             & TF-DCon         & 0.3071 & 0.3691 & 0.4377 & 0.6150  \\ \cmidrule{2-7} 
                           & \multirow{2}{*}{NRMS}       & $\text{TF-DCon}_{RI}$ & 0.2916 & 0.3479 & 0.4010  & 0.589  \\
                           &                             & TF-DCon         & 0.2997 & 0.3597 & 0.4279 & 0.6000    \\ \cmidrule{2-7} 
                           & \multirow{2}{*}{Fastformer} & $\text{TF-DCon}_{RI}$ & 0.2898 & 0.3589 & 0.4197 & 0.5938 \\
                           &                             & TF-DCon         & 0.3022 & 0.3637 & 0.4334 & 0.6096 \\ \midrule
\multirow{6}{*}{Goodreads} & \multirow{2}{*}{NAML}       & $\text{TF-DCon}_{RI}$ & 0.5257 & 0.7988 & 0.4626 & 0.9982 \\
                           &                             & TF-DCon         & 0.5411 & 0.8033 & 0.4704 & 0.9984 \\ \cmidrule{2-7} 
                           & \multirow{2}{*}{NRMS}       & $\text{TF-DCon}_{RI}$ & 0.5294 & 0.7993 & 0.4679 & 0.9986 \\
                           &                             & TF-DCon         & 0.5453 & 0.8054 & 0.4751 & 0.9985 \\ \cmidrule{2-7} 
                           & \multirow{2}{*}{Fastformer} & $\text{TF-DCon}_{RI}$ & 0.5496 & 0.8055 & 0.4602 & 0.9989 \\
                           &                             & TF-DCon         & 0.5548 & 0.8092 & 0.4851 & 0.9986 \\ \midrule
\multirow{6}{*}{MovieLens} & \multirow{2}{*}{NAML}       & $\text{TF-DCon}_{RI}$ & 0.8361 & 0.8402 & 0.1813 & 0.7432 \\
                           &                             & TF-DCon         & 0.8484 & 0.8494 & 0.1873 & 0.7475 \\ \cmidrule{2-7} 
                           & \multirow{2}{*}{NRMS}       & $\text{TF-DCon}_{RI}$ & 0.8126 & 0.8344 & 0.1764 & 0.7340  \\
                           &                             & TF-DCon         & 0.8149 & 0.8371 & 0.1798 & 0.7446 \\ \cmidrule{2-7} 
                           & \multirow{2}{*}{Fastformer} & $\text{TF-DCon}_{RI}$ & 0.8085 & 0.8352 & 0.1821 & 0.7437 \\
                           &                             & TF-DCon         & 0.8251 & 0.8429 & 0.1836 & 0.7465 \\ \bottomrule
\end{tabular}
}
\label{tab:ablation-user-interest}
\end{table}
\subsubsection{Moderation Check.} To check whether the condensed dataset contain toxic contents, we conduct moderation check via the endpoints provided by OpenAI\footnote{https://platform.openai.com/docs/guides/moderation}. Specifically, we randomly sample 100 items from each dataset to assess their contents. The results are presented in table~\ref{tab:moderations}, where the "{Mean}" signifies the mean {violation} score of the contents in items encompassing the corresponding toxic words and "Var" denotes the associated variance. {The {violation} score represents the model's confidence in input violations of the OpenAI category policy. This value ranges between 0 and 1, with higher values indicating higher confidence that the input contains the corresponding toxic words.} We can observe from the table that the condensed datasets derived from the three original datasets do not incorporate any of the toxic contents.
\begin{table}[]
\vskip -0.1in
\caption{Moderation Check.}
\begin{tabular}{l|ll|ll|ll}
\toprule
\multicolumn{1}{l|}{{Dataset}} & \multicolumn{2}{l|}{{MIND}}                                    & \multicolumn{2}{l|}{{Goodreads}}                                   & \multicolumn{2}{l}{{MovieLens}}                                    \\ \midrule
{{Violation Category}}                  & \multicolumn{1}{l}{{Mean}} & \multicolumn{1}{l|}{{Var}} & \multicolumn{1}{l}{{Mean}} & \multicolumn{1}{l|}{{Var}} & \multicolumn{1}{l}{{Mean}} & \multicolumn{1}{l}{{Var}} \\ \midrule
sexual                                & 4.94E-03                          & 2.13E-03                          & 8.45E-04                          & 1.43E-05                          & 3.08E-04                          & 1.13E-06                         \\
hate                                  & 6.72E-05                          & 3.97E-08                          & 3.33E-05                          & 1.90E-08                          & 2.11E-05                          & 3.75E-09                         \\
harassment                            & 1.57E-04                          & 2.64E-07                          & 4.55E-05                          & 6.64E-09                          & 4.64E-05                          & 4.29E-08                         \\
self-harm                             & 2.46E-06                          & 8.65E-11                          & 3.04E-06                          & 1.77E-10                          & 2.71E-06                          & 8.68E-11                         \\
sexual/minors                         & 1.05E-05                          & 4.26E-09                          & 1.55E-05                          & 1.09E-08                          & 2.68E-06                          & 5.28E-11                         \\
hate/threatening                      & 1.18E-06                          & 2.12E-11                          & 1.01E-06                          & 1.70E-11                          & 1.11E-06                          & 2.10E-11                         \\
violence/graphic                      & 4.77E-04                          & 4.31E-06                          & 7.49E-05                          & 9.46E-08                          & 5.03E-05                          & 3.11E-08                         \\
self-harm/intent                      & 2.43E-07                          & 1.43E-12                          & 7.80E-08                          & 1.18E-13                          & 1.38E-07                          & 3.61E-13                         \\
self-harm/instructions                & 7.96E-07                          & 6.16E-12                          & 3.07E-08                          & 3.02E-14                          & 9.53E-08                          & 2.98E-13                         \\
harassment/threatening                & 2.13E-05                          & 1.36E-08                          & 1.75E-06                          & 2.60E-11                          & 4.18E-06                          & 4.96E-10                         \\
violence                              & 3.40E-03                          & 2.57E-04                          & 4.79E-03                          & 3.50E-04                          & 3.42E-03                          & 1.10E-04                         \\ \bottomrule
\end{tabular}
\label{tab:moderations}
\vskip -0.1in
\end{table}

\subsection{Case Study}
In this section, we conduct case study to examine the reasonability of the condensed data by investing the correlation between the history of the synthesized users and their corresponding real users. The following example is randomly sampled from the MovieLens dataset. We show the history of the synthesized user and three related real users. In the following, the history of synthesized user consists of short introduction of the interacted movies while the histories of real users consist of the title and description of the interacted movies. We can observe that the preference information of those real users can be roughly covered by this synthesized user, which may be the reason why models trained on the condensed data since it provides adequate information for sufficient model training.

\begin{mybox3}{Synthesized User}
\textit{\textbf{Interaction History}}:
\begin{itemize}[leftmargin=*,itemsep=0pt,topsep=0pt,parsep=0pt]
\item \textit{My Best Friend's Wedding is about a romantic comedy.} \gray{from: \textbf{//Real User 2}}
\item \textit{Back to the Future is about a high school student who time travels to the past and must ensure his parents fall in love to preserve his own existence.} \gray{from: \textbf{//Real User 2}}
\item \textit{Grosse Pointe Blank is about a hitman facing unexpected dilemmas at a high school reunion.} \gray{from: \textbf{//Real User 3}}
\item \textit{"Robin Hood: Men in Tights" is about a skilled archer leading a funny rebellion against a corrupt ruler, with the help of his misfit Merry Men.} \gray{from: \textbf{//Real User 1 \& Real User 2 \& Real User 3}}
\item \textit{Mad City is about a hostage situation in a museum, delving into the power struggle between a security guard and a journalist, while examining media ethics and the blurred boundaries of truth.} \gray{from: \textbf{//Real User 1}}
\item \textit{The Wizard of Oz is about Dorothy's journey in the magical land of Oz, where she seeks the help of the Great Wizard to return home, facing challenges and learning valuable lessons along the way.} \gray{from: \textbf{//Real User 1}}
\item \textit{This Is Spinal Tap is a mockumentary comedy released in 1984.} \gray{from: \textbf{//Real User 1}}
\end{itemize}
\end{mybox3}

\begin{mybox3}{Real User 1}
\textit{\textbf{Interaction History}}:
\begin{itemize}[leftmargin=*,itemsep=0pt,topsep=0pt,parsep=0pt]
    \item \textit{Snow White and the Seven Dwarfs (1937)}, with description:\begin{itemize}[leftmargin=*,itemsep=0pt,topsep=0pt,parsep=0pt]
        \item \textit{"Snow White and the Seven Dwarfs is a timeless animated film released in 1937. It tells the story of a young princess named Snow White who finds herself in the care of seven lovable dwarfs after escaping from her evil stepmother. With its enchanting tale, memorable songs, and captivating animation, this classic movie continues to captivate audiences of all ages."}
    \end{itemize}
    \item \textit{Robin Hood: Men in Tights (1993)}, with description:\begin{itemize}[leftmargin=*,itemsep=0pt,topsep=0pt,parsep=0pt]
        \item \textit{"Robin Hood: Men in Tights (1993), a comedic retelling of the classic Robin Hood tale, a skilled archer named Robin must outwit the corrupt Prince John and win the heart of the beautiful Maid Marian. With a band of misfit Merry Men by his side, Robin leads a hilarious and daring rebellion against the unjust ruler, using wit, skill, and plenty of laughs along the way."}
    \end{itemize}
    \item \textit{The Wizard of Oz (1939)}, with description:\begin{itemize}[leftmargin=*,itemsep=0pt,topsep=0pt,parsep=0pt]
        \item \textit{"The Wizard of Oz is a classic fantasy film released in 1939. It tells the story of a young girl named Dorothy, who is transported to the magical land of Oz after a tornado hits her Kansas farm. In Oz, Dorothy encounters various colorful characters, including the Scarecrow, the Tin Man, and the Cowardly Lion, as she embarks on a journey to find the Great Wizard who can help her return home. Along the way, they face challenges from the Wicked Witch of the West, and learn important lessons about friendship, courage, and the importance of following your dreams. The Wizard of Oz is beloved for its stunning visuals, memorable songs, and timeless themes."}
    \end{itemize}
    \item \textit{Mad City (1997)}, with description:\begin{itemize}[leftmargin=*,itemsep=0pt,topsep=0pt,parsep=0pt]
        \item \textit{"Mad City, a 1997 film, centers on a hostage crisis in a museum, delving into the power struggle between a disgruntled security guard and a manipulative journalist, each vying to control the crisis narrative. Through intense dialogue and suspense, Mad City critically examines media ethics and the blurred boundaries between truth and fiction."}
    \end{itemize}
    \item \textit{This Is Spinal Tap (1984)}, with description:\begin{itemize}[leftmargin=*,itemsep=0pt,topsep=0pt,parsep=0pt]
        \item \textit{"This Is Spinal Tap is a comedy mockumentary film released in 1984."}
    \end{itemize}
\end{itemize}
\end{mybox3}

\begin{mybox3}{Real User 2}
\textit{\textbf{Interaction History}}:
\begin{itemize}[leftmargin=*,itemsep=0pt,topsep=0pt,parsep=0pt]
    \item \textit{My Best Friend's Wedding (1997)}, with description: \begin{itemize}[leftmargin=*,itemsep=0pt,topsep=0pt,parsep=0pt]
        \item \textit{"My Best Friend's Wedding is a romantic comedy film released in 1997."}
    \end{itemize}
    \item \textit{Snow White and the Seven Dwarfs (1937)}, with description:\begin{itemize}[leftmargin=*,itemsep=0pt,topsep=0pt,parsep=0pt]
        \item \textit{"Snow White and the Seven Dwarfs is a timeless animated film released in 1937. It tells the story of a young princess named Snow White who finds herself in the care of seven lovable dwarfs after escaping from her evil stepmother. With its enchanting tale, memorable songs, and captivating animation, this classic movie continues to captivate audiences of all ages."}
    \end{itemize}
    \item \textit{Die Hard (1988)}, with description:\begin{itemize}[leftmargin=*,itemsep=0pt,topsep=0pt,parsep=0pt]
        \item \textit{ "Die Hard is an action film released in 1988."}
    \end{itemize}
    \item \textit{Robin Hood: Men in Tights (1993)}, with description:\begin{itemize}[leftmargin=*,itemsep=0pt,topsep=0pt,parsep=0pt]
        \item \textit{"In Robin Hood: Men in Tights (1993), a comedic retelling of the classic Robin Hood tale, a skilled archer named Robin must outwit the corrupt Prince John and win the heart of the beautiful Maid Marian. With a band of misfit Merry Men by his side, Robin leads a hilarious and daring rebellion against the unjust ruler, using wit, skill, and plenty of laughs along the way."}
    \end{itemize}
    \item \textit{{Back to the Future (1985)}}, with description:\begin{itemize}[leftmargin=*,itemsep=0pt,topsep=0pt,parsep=0pt]
        \item \textit{"Back to the Future is a science fiction film released in 1985. The movie follows the story of a high school student named Marty McFly who, with the help of a eccentric scientist, travels back in time to the year 1955. Marty must find a way to return to the present and also ensure that his parents meet and fall in love, or else he risks erasing his own existence. The film combines comedy, adventure, and a touch of romance as it explores the consequences of altering the past and the importance of family."}
    \end{itemize}
\end{itemize}
\end{mybox3}

\begin{mybox3}{Real User 3}
\textit{\textbf{Interaction History}}:
\begin{itemize}[leftmargin=*,itemsep=0pt,topsep=0pt,parsep=0pt]
    \item \textit{Robin Hood: Men in Tights (1993)}, with description:\begin{itemize}[leftmargin=*,itemsep=0pt,topsep=0pt,parsep=0pt]
        \item \textit{"Robin Hood: Men in Tights" (1993), a comedic retelling of the classic Robin Hood tale, a skilled archer named Robin must outwit the corrupt Prince John and win the heart of the beautiful Maid Marian. With a band of misfit Merry Men by his side, Robin leads a hilarious and daring rebellion against the unjust ruler, using wit, skill, and plenty of laughs along the way."}
    \end{itemize}
    \item \textit{Chasing Amy (1997)}, with description:\begin{itemize}[leftmargin=*,itemsep=0pt,topsep=0pt,parsep=0pt]
        \item \textit{"‘Chasing Amy’ is a 1997 romantic comedy-drama film."}
    \end{itemize}
    \item \textit{In the Name of the Father (1993)}, with description:\begin{itemize}[leftmargin=*,itemsep=0pt,topsep=0pt,parsep=0pt]
        \item \textit{"In the Name of the Father is a gripping drama film released in 1993. It tells the story of Gerry Conlon, a young Irishman wrongfully convicted of an IRA bombing. Set in the backdrop of the Troubles in Northern Ireland, the film explores themes of injustice, incarceration, and the quest for truth and justice. Directed by Jim Sheridan and featuring remarkable performances by Daniel Day-Lewis and Pete Postlethwaite, this emotional and thought-provoking movie delves into the complexities of the legal system and the enduring power of family bonds."}
    \end{itemize}
    \item \textit{Nightmare on Elm Street, A (1984)}, with description:\begin{itemize}[leftmargin=*,itemsep=0pt,topsep=0pt,parsep=0pt]
        \item \textit{"A Nightmare on Elm Street is a horror film released in 1984. Directed by Wes Craven, the movie tells the story of a group of teenagers who are haunted in their dreams by a supernatural killer named Freddy Krueger. As they struggle to stay awake and escape his clutches, they soon realize that death in their dreams also means death in reality. Filled with suspenseful moments and unnerving visual effects, A Nightmare on Elm Street delivers a chilling and unforgettable experience for horror enthusiasts."}
    \end{itemize}
    \item \textit{Grosse Pointe Blank (1997)}, with description:\begin{itemize}[leftmargin=*,itemsep=0pt,topsep=0pt,parsep=0pt]
        \item \textit{"Grosse Pointe Blank is a 1997 movie that revolves around a skilled and professional hitman who, during a high school reunion, unexpectedly finds himself confronting personal and moral dilemmas."}
    \end{itemize}
\end{itemize}
\end{mybox3}
\subsection{{Implications}}
{The results show that that our proposed method is effective in condense content-based recommendation datasets, achieving comparable model training performance with even $5\%$ data. Our method designs a novel condensing paradigm, training-free, from previous iterative-optimization based ones. Benefiting from this design, our method is significantly more efficient. Besides, we enable the generation of discrete texts via ChatGPT and preserve users' preference via user-level condensation module. Therefore, incorporating our method to condense datasets for daily model training can save lots of resources costs without performance degradation in practice.}

\section{{Limitations}}
\label{sec:limitations}
{Although TF-DCon has been introduced to address the issue of resources demanding during model training by condensing datasets, it introduces new limitations: (1) Although the condensation process is efficient, it requires a pre-process of the dataset by ChatGPT/other LLMs, the efficiency of which can be unstable when we utilize ChatGPT for the pre-process since it is affected by the network latency. (2) Since the cost of calling ChatGPT is linear to the size of dataset, we may need to train the models based on the condensed datasets for a certain times to achieve the cost-gain balance. (3) As shown in the experiments, the performance of TF-DCon with open source LLMs is still limited.}
\section{Conclusion}
\label{sec:conclusion}
Advanced content-based recommendation models rely on training with large datasets, which is costly. To solve this problem, in this paper, we investigate how to condense the dataset of content-based recommendation into a small synthetic dataset, enabling models to achieve performance comparable to those trained on full datasets. We propose a novel method TF-DCon where a prompt-evolving module is proposed to adapt ChatGPT to condense contents of items, and users and their corresponding historical interactions are condensed via a curated clustering-based synthesis module. This work represents the first exploration of dataset condensation for content-based recommendation and the first non-iterative synthetic data optimization approach. The experimental results on multiple real-world datasets verify the effectiveness of our proposed methods. Notably, we approximate up to $97\%$ of the original performance while reducing the dataset size by $95\%$ (i.e., on the MIND dataset). The condensation process of TF-DCon is significantly faster than the method that is based on nested optimization. 


\section*{CRediT Authorship Contribution Statement}
    \textbf{Jiahao Wu}: Conceptualization, Methodology, Partial experiments, Validation, Formal analysis, Investigation, Manuscript writing. \textbf{Qijiong Liu}: Data Curation, Main experiments, Validation, Formal analysis, Investigation. \textbf{Hengchang Hu}: Formal Analysis, Review, Editing. \textbf{Wenqi Fan}: Conceptualization, Formal Analysis, Review, Supervision. \textbf{Shengcai Liu}: Formal Analysis, Review, Editing, Supervision. \textbf{Qing Li}: Conceptualization, Formal Analysis, Review, Supervision. \textbf{Xiaoming Wu}: Conceptualization, Formal Analysis, Review, Supervision. \textbf{Ke Tang}: Conceptualization, Formal Analysis, Review, Editing, Supervision.
\section*{Resource Availability}
    Code and data will be made available on request.

\section*{Declaration of Generative AI and AI-assisted Technologies in the Writing Process}

During the preparation of this work the author(s) used ChatGPT in order to to improve language and readability, with caution. After using this tool/service, the author(s) reviewed and edited the content as needed and take(s) full responsibility for the content of the publication.

\bibliographystyle{cas-model2-names}
\bibliography{references}
\end{document}